\documentclass{article} 
\usepackage[T1]{fontenc}
\usepackage{iclr2027_conference,times}

\usepackage{amsmath,amsfonts,bm}

\def\eqref#1{equation~\ref{#1}}

\def\1{\bm{1}}

\DeclareMathAlphabet{\mathsfit}{\encodingdefault}{\sfdefault}{m}{sl}
\SetMathAlphabet{\mathsfit}{bold}{\encodingdefault}{\sfdefault}{bx}{n}

\usepackage{hyperref}
\hypersetup{colorlinks=false, citebordercolor={0 1 0},
  linkbordercolor={1 0 0}, urlbordercolor={0 1 1}, pdfborder={0 0 1}}
\usepackage{url}
\usepackage{graphicx}
\usepackage{booktabs}
\usepackage{capt-of}
\usepackage{array}
\usepackage{multirow}
\usepackage{float}
\usepackage{wrapfig}
\usepackage{enumitem}
\usepackage{tcolorbox}
\title{TRACE: Governing Memory Validity\\
in Evolving Multi-Agent Systems}

\author{\begin{minipage}[t]{\dimexpr\textwidth-2\tabcolsep\relax}
\raggedright
\textbf{Wenjun Xiong\textsuperscript{1,4,\ensuremath{\dagger}}, Shengtao Zhang\textsuperscript{1},
Shangding Gu\textsuperscript{3}, Bo Tang\textsuperscript{4},}\\
\textbf{Zhiyu Li\textsuperscript{4}, Feiyu Xiong\textsuperscript{4},
Ying Wen\textsuperscript{1,2,\ensuremath{*}},
Muning Wen\textsuperscript{1,\ensuremath{*}}}\\
\normalfont
\textsuperscript{1}Shanghai Jiao Tong University\enspace
\textsuperscript{2}Shanghai Innovation Institute\\
\textsuperscript{3}UC Berkeley\enspace
\textsuperscript{4}MemTensor(Shanghai) Technology Co., Ltd.\\
\texttt{\{xiongwenjun,ying.wen,muningwen\}@sjtu.edu.cn}
\end{minipage}}

\iclrfinalcopy
\begin{document}

\maketitle
\lhead{Preprint}
\begingroup
\renewcommand{\thefootnote}{\fnsymbol{footnote}}
\footnotetext[1]{Corresponding authors.}\footnotetext[2]{Work done during internship at MemTensor.}
\endgroup

\begin{abstract}
  Persistent memory lets language-model agents carry information across
  long-running collaborations, but leaves a lifecycle question open: what may a
  returning agent still act on once the shared state has changed? A memory can
  be correctly retrieved, relevant to the current task, and faithful to its
  source, and nonetheless be inadmissible for action: an itinerary saved before
  a pause still names the hotel the team has since
  replaced~(Figure~\ref{fig:return-dilemma}).    We formalize this as
  \emph{temporal memory admission} and present TRACE, a training-free layer that
  treats re-entry as an eligibility decision rather than a storage or retrieval
  operation, reconciling a departure checkpoint against absence-period updates,
  resolving explicit and implicit invalidation, and releasing a bounded Return
  View only when it covers the returning role's open obligations. We evaluate
  TRACE under three actor models on Memora, STALE Type~II, and a derived
  ManBench-Return setting, each recast as return episodes: one agent departs,
  four teammates change the shared state, and the agent rejoins. What separates
  methods is not overall accuracy but whether one can retain valid memory and
  reject stale memory at once, and no single-policy baseline can: \textsc{Restore}
  (reinstate the departure checkpoint in full) admits stale state, \textsc{Reset} (start
  the return from an empty memory) discards valid state, each bottoming out at
  0\% on one of the two. TRACE is the only method high on both, reaching 92.6--98.3\%
  valid-information availability with 98.4--99.5\% invalid-information rejection
  on ManBench-Return, within 3.8 points of the best baseline's overall accuracy.
  On STALE Type~II it improves Overall over the strongest comparison policy by
  22.3 (Qwen), 18.5 (Gemini), and 27.5 (DeepSeek) points at roughly $2.3\times$
  their tokens, while a write-time consolidation
  pipeline is more accurate still at $3.99\times$ TRACE's\footnote{Code is available at \url{https://github.com/xiong-wenjun/TRACE}.}. 
\end{abstract}

\section{Introduction}
\label{sec:introduction}

\begin{figure}[t]
  \centering
  \setlength{\parskip}{0pt}
  \setlength{\abovecaptionskip}{-6pt}
  \includegraphics[width=\linewidth]{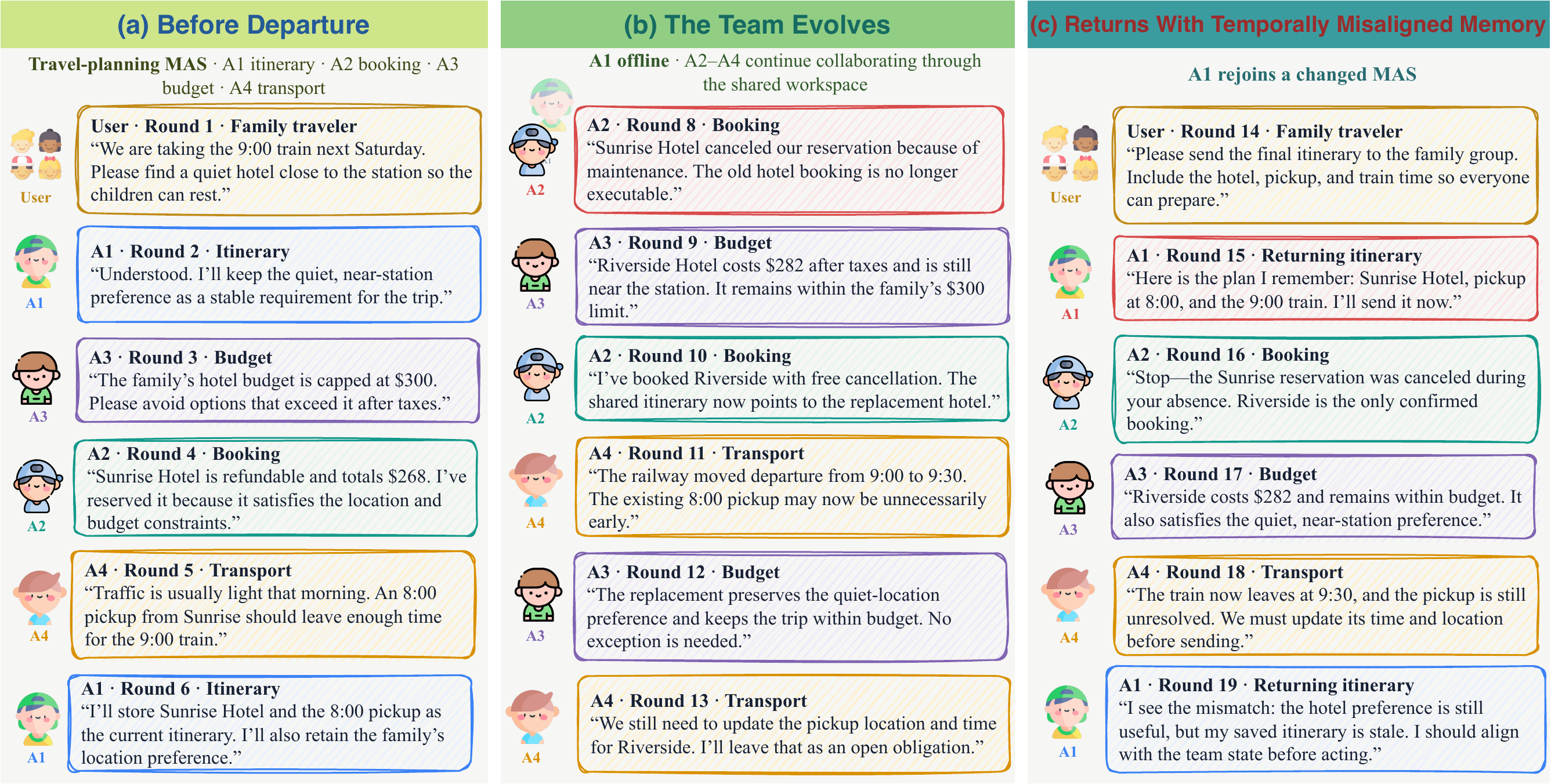}
  \caption{The agent-return dilemma under context evolution. A1 leaves with a
  coherent travel plan. During its absence, the team replaces the hotel booking,
  changes the train schedule, and leaves the pickup time unresolved. When A1
  returns, its stored itinerary is still relevant and was correct when recorded,
  yet parts of it are no longer admissible for action.}
  \label{fig:return-dilemma}
\end{figure}

Long-running multi-agent systems do not keep a fixed roster: agents pause, are
reassigned, and rejoin a workspace the remaining members have continued to change.
Figure~\ref{fig:return-dilemma} introduces the resulting difficulty. Agent A1 departs with a
coherent itinerary; during its absence the team replaces the hotel, the railway moves
the departure time, and the pickup is left as an open obligation. A1's stored itinerary
is still topically relevant and was accurate when recorded, yet parts of it can no
longer be acted on. Faithful retrieval does not establish current eligibility for
action. The example poses the question this paper studies: at an authenticated return,
which parts of a retained memory remain admissible, and on what basis is that decided?

Persistent memory and retrieval systems are designed to reuse historical information
\citep{park2023generative, shinn2023reflexion, lewis2020rag, packer2023memgpt,
chhikara2025mem0, maharana2024locomo, wu2025longmemeval}. Memora evaluates selective forgetting \citep{uddin2026memora},
while temporal knowledge graphs and transactional belief protocols track facts that
evolve \mbox{\citep{rasmussen2025zep, li2026memtx}}. Each improves what a store retains or
retrieves. Recent memory managers add reflection and temporal organization
\citep{agentmemsurvey2024,kang-etal-2025-memory,tan-etal-2025-prospect}, and newer
benchmarks separate long-term recall from continual repair
\citep{tan-etal-2025-membench,memoryagentbench2026,jia-etal-2025-evaluating}; none
decides what a particular agent may act on when it rejoins a changed
team. Nor does the failure require an adversary \citep{chen2024agentpoison,
xiong2026mapleguard}: ordinary task evolution can strand records that were trustworthy when
written. Two controlled findings confirm the gap. On STALE, LightMem fails implicit
policy adaptation in 78.6\% of cases in which the updated evidence had already been
retrieved \citep{chao2026stale}. On ManBench, Qwen3-235B answers 56.85\% of initially
correct questions incorrectly after misleading group discussion and memory
consolidation \citep{xu2026manbench}.

Prior work separates storage and retrieval, belief maintenance, and social
propagation~\citep{doyle1979truth,zhong2024memorybank,xu2026manbench}.
The return event remains less specified: a record may be retrievable and relevant,
yet fail the returning role's current obligations.

We formalize this check as \emph{temporal memory admission}: the returning agent
reassesses retained memory against current state, identity, task purpose, and open
obligations. Recency alone is insufficient: old constraints may still bind, while
recent records may lack authorization, provenance, or required state.

TRACE (\textbf{T}emporal \textbf{R}eturn \textbf{A}dmission under \textbf{C}ontext
\textbf{E}volution) enforces this contract at return as a training-free layer between
a memory backend and the agent. It reconciles the departure checkpoint with the
absence-period delta, resolves explicit invalidation, and verifies candidate implicit
dependencies. Surviving records pass authorization, temporal, applicability, and
provenance gates before a budgeted selector covers open obligations. The signed Return
View is released only after validation and coverage, and private experience is readmitted
only with its bound work-state item.

We evaluate TRACE under a common five-agent return protocol on three settings that
isolate distinct forms of state change: explicit replacement in Memora, implicit
contextual invalidation in STALE Type~II, and propagated conflict in ManBench-Return,
which we derive from ManBench conflict scenarios and report separately from the original
benchmark. Each setting is run under three actor models, and the protocol also delimits
where the layer is unnecessary. 

On Memora, where replacement information is explicit,
ordinary storage semantics already suffice: TRACE offers no consistent gain and trails a
\textsc{Reset} policy on overall accuracy. On STALE Type~II it raises overall accuracy over
the strongest comparison policy by 22.3 points with Qwen, 18.5 with Gemini, and 27.5
with DeepSeek, and with Qwen it raises
Lifecycle Success from 18.0\% to 39.0\%, at roughly 2.3 times the tokens per episode of
the cheaper lifecycle policies; the write-time consolidation baseline \textsc{CUPMem} remains more
accurate still on every reported score, at 3.99 times TRACE's token total. On
ManBench-Return it reaches 92.6--98.3\% valid-information availability with 98.4--99.5\%
invalid-information rejection, within 3.8 points of the best single-policy baseline's
overall accuracy, whereas \textsc{Reset} rejects valid and stale state alike and \textsc{Restore} preserves
both. Return-time admission therefore trades bounded overhead for joint preservation and
rejection that neither policy provides.

Our contributions are threefold:
\begin{itemize}[leftmargin=*]
  \item \textbf{Formalizing temporal memory admission.}
  We formalize the conditions under which retained memories remain admissible for action when an agent rejoins an evolving team. This formulation separates historical correctness and query relevance from current admissibility under explicit and implicit state changes.

  \item \textbf{Auditable admission at agent return.}
  We introduce TRACE, a training-free layer that combines source-verified implicit invalidation with item-level eligibility checks and obligation-aware memory selection. TRACE assembles an auditable Return View, releasing it only when validation and obligation coverage succeed, while readmitting private experience only if its bound work-state item remains eligible.

  \item \textbf{Empirical benefits and applicability.}
  Across three return benchmarks and three actor models, we show that TRACE jointly preserves valid information and rejects stale information, with gains concentrated in implicit and propagated invalidation and limited benefit under explicit updates. Four mechanism ablations, evaluations with 4--32 active agents, and tests across four memory backends further characterize its behavior and applicability.
\end{itemize}
    

\section{Temporal Memory Admission}
\label{sec:problem}

\subsection{Lifecycle boundary}

Let agent $a$ depart from team $\mathcal{A}$ at time $t_d$ and return at
$t_r > t_d$, preserving its principal and role but starting a fresh workload
instance. Over the absence interval the shared work state evolves from $C_{t_d}$
to $C_{t_r}$ according to a recorded delta $\Delta_{(t_d,t_r]}$, while $a$ retains
a pre-departure bank $M_a^-$ of shared work-state records and private episodic
experience. A return mechanism must expose enough of $M_a^-$ to preserve useful
continuity without making superseded or unauthorized premises executable again.

A work-state item $m$ carries content, a source receipt, an epoch and validity
interval, principal and role scope, task purpose, dependency links, and
supersession status, where an epoch indexes one recorded revision of the shared
state. Relevance to a fresh query $q$ is not sufficient for admissibility: there
exist $m$ and $q$ such that
\begin{equation}
\mathrm{Rel}(m,q) = 1
\quad\text{and}\quad
\mathrm{Adm}\!\left(m \mid a,\, C_{t_r},\, \Delta_{(t_d,t_r]},\, O_{t_r}\right) = 0,
\end{equation}
where $O_{t_r}$ denotes the obligations of the returning role. This second
predicate, which turns on authority and state dependence rather than on similarity
or recency, is the decision studied in this paper.

\subsection{Two temporal failure modes}

A valid item loses eligibility in two ways. Explicit, linkable invalidation
supplies a machine-readable relation: a revocation, a replacement edge, a closed
validity interval, or a same-identity update. A conventional store resolves it
whenever the relation is preserved \citep{rasmussen2025zep, yadav2026memstrata}.
Implicit invalidation (II) instead changes a premise the item depends on without
declaring the item false, as when a changed destination invalidates a train
recommendation that no later event mentions. Resolving II requires reasoning over
the item's current use together with the absence-period delta, the regime that
STALE Type~II isolates \citep{chao2026stale}.

Source quality and cross-agent propagation cut across this distinction rather than
extending it. A stale item may carry an authentic receipt and a current item an
unverified one, and repeated transmission can make wrong state appear well
supported \citep{xu2026manbench} without creating a new temporal failure mode. We
therefore treat provenance, state dependence, and temporal validity as separate
gates.

\subsection{Desired contract}

A correct return mechanism should preserve the valid information the fresh workload
needs, reject items that no longer support current action, and fail closed when the
available evidence cannot cover critical obligations. The first two requirements
define Valid Information Availability and Invalid Information Rejection
(Section~\ref{sec:metrics}); the third is enforced by the coverage gate
(Section~\ref{sec:cover}). Together they exclude both unconditional restoration and
unconditional reset, and they motivate a bounded, inspectable artifact at the
lifecycle boundary in place of unrestricted access to $M_a^-$, following the
fail-safe-default principle of access-control design \citep{saltzer1975protection}.

\section{TRACE}
\label{sec:method}
\begin{figure}[t]
  \centering
  \setlength{\parskip}{0pt}
  \setlength{\abovecaptionskip}{-6pt}
  \includegraphics[width=\linewidth,trim=0 24 0 24,clip]{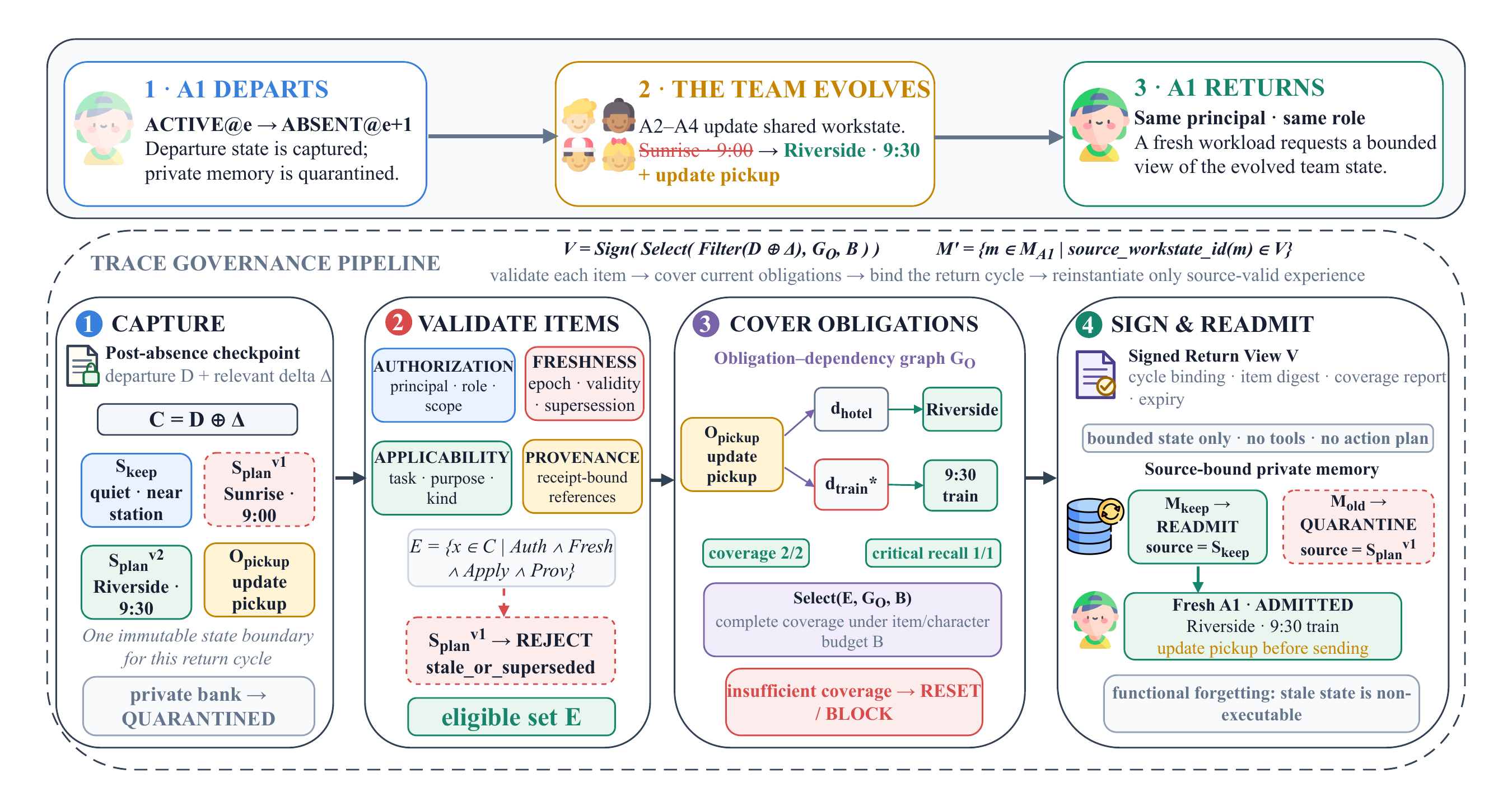}
  \caption{TRACE at an authenticated return boundary. The system captures one
  immutable departure-and-delta view, validates candidate items, selects a
  budgeted set that covers current obligations, and signs a Return View. Private
  experience is readmitted only when its bound source state survives.}
  \label{fig:trace-overview}
\end{figure}

Figure~\ref{fig:trace-overview} summarizes the four stages of the layer. TRACE establishes one
  immutable state boundary per return cycle and governs which records become visible to
  the fresh workload. Appendix~\ref{app:method-details} gives formal definitions and complete
return examples (Appendices~\ref{app:return-case} and~\ref{app:benchmark-case}).

\subsection{Capture a return-cycle checkpoint}

At departure, TRACE records a checkpoint $D_a$: the portion of $C_{t_d}$ visible to $a$
under its principal, role, and task scope, together with the source receipts and state
epoch that make it auditable. The private episodic bank is quarantined rather than copied
forward. At return, TRACE binds $D_a$ to the relevant absence delta and materializes the
post-absence checkpoint $C = D_a \oplus \Delta$, a single state against which all
subsequent decisions are made.

The candidate pool $\mathcal{P}$ drawn from $C$ contains verified facts, progress
frontiers, open obligations, and role and interface constraints. Private experiences stay
in the agent's own bank, each bound to the work-state item that sourced it. Candidate
construction is deliberately broader than final admission: omitting a useful record is
irreversible at this stage, whereas the later gates can still reject a false positive.

\subsection{Resolve explicit and implicit invalidation}

TRACE first applies direct lifecycle evidence, removing an item whose source epoch
precedes an incompatible target epoch, whose validity interval does not cover $t_r$, or
which a recorded supersession or revocation link targets. Updates that carry no such link
require the two-stage test that distinguishes II. A high-recall scan proposes
absence-period events that may change a premise of the candidate's current use. A grounded
verifier then receives each source-bound pair and decides whether the later event
invalidates that use in the returning task, deciding on checkpoint evidence alone rather
than on benchmark annotations. Separating proposal from verification confines the
expensive pairwise judgment to plausible dependencies and keeps topical similarity from
being read as contradiction.

\subsection{Validate eligible items}

Every surviving candidate passes four item-level gates:
\begin{equation}
E = \{\, m \in \mathcal{P} : A(m) \wedge T(m) \wedge U(m) \wedge P(m) \,\},
\label{eq:eligible}
\end{equation}
where $A$ checks authorization for the principal, role, and scope; $T$ checks epoch,
validity, and supersession; $U$ checks applicability to the task, purpose, and item kind;
and $P$ requires bound receipts for referenced source state. Authorization and
applicability follow an attribute-based policy view \citep{hu2014abac}, while $P$ records
source derivation \citep{moreau2013prov} without certifying that the source statement is
true. Each item must also fit an individual size budget. Keeping the predicates separate
yields a reason-coded audit trail and prevents a recent but unauthorized record from
passing as fresh.

\subsection{Cover obligations under a context budget}
\label{sec:cover}

Admission is a set decision as well as an item decision. Let the graph $G_O$ link current
obligations and progress frontiers to the eligible records supporting them. TRACE first
admits the mandatory open obligations and frontiers, then greedily adds low-cost records
that cover uncovered required and critical dependencies, approximating
\begin{equation}
\min_{S \subseteq E} \sum_{m \in S} c(m)
\quad \text{s.t.} \quad
\mathrm{Cov}(S) \ge \gamma, \;\;
\mathrm{Crit}(S) \ge \rho, \;\;
|S| \le B_n, \;\;
\sum_{m \in S} \ell(m) \le B_c,
\end{equation}
where $c(m)$ is disclosure cost, $\ell(m)$ is text length, and $B_n$ and $B_c$ bound item
count and characters. When no selection meets the required coverage, the return is reset
or blocked rather than issued with an obligation silently omitted. The objective is
related to weighted set cover \citep{chvatal1979greedy}; the hard budgets and mandatory
items make our selector a constrained heuristic for which we claim no classical
approximation guarantee. The explicit context budget also reflects the uneven use of evidence across long
contexts~\citep{liu2024lostmiddle}.

\subsection{Sign and readmit a bounded Return View}

The selected set is serialized as a Return View carrying its return-cycle identifier,
principal, role, task, epoch, expiry, item digests, and a coverage report. Before
exposure, TRACE recomputes the digest and rechecks the event, identity, epoch, expiry,
and item predicates. The view supplies bounded state, not tools or an action plan. A
private experience is reinstated only if the work-state item it was bound to remains in
the signed view. This realizes functional forgetting: rejected state may persist in an
audit store, yet it cannot serve as a premise for the returning workload.

\section{Experiments}
\label{sec:experiments}

\subsection{Research questions and settings}

We ask when return-time admission improves validity and at what token cost (RQ1),
whether it preserves valid state while rejecting propagated errors (RQ2), and whether it
remains stable across team sizes and memory backends (RQ3). The settings are Memora for
longitudinal retention and forgetting~\citep{uddin2026memora}, STALE Type~II for implicit
contextual invalidation~\citep{chao2026stale}, and ManBench-Return for propagated conflict
and state admission~\citep{xu2026manbench}.
\begin{table}[t]
\caption{Main comparison across three return benchmarks (\%). Overall denotes
benchmark-specific overall accuracy (Section~\ref{sec:metrics}). Memora averages
the three departure conditions; ManBench-Return averages Old-Valid and
Old-Stale using unrounded condition scores. STALE aggregation follows Appendix~\ref{app:stale-settings}.
Bold marks the best value within each actor and metric.}
\label{tab:main-results}
\centering
\fontsize{8}{9.5}\selectfont
\setlength{\tabcolsep}{1.5pt}
\begin{tabular*}{\linewidth}{@{\extracolsep{\fill}}llccccccccc@{}}
\toprule
& & \multicolumn{3}{c}{Memora} & \multicolumn{3}{c}{STALE Type~II} & \multicolumn{3}{c}{ManBench-Return} \\
\cmidrule(lr){3-5}\cmidrule(lr){6-8}\cmidrule(lr){9-11}
Model & Method & Overall$\uparrow$ & VIA$\uparrow$ & IIR$\uparrow$ & Overall$\uparrow$ & VIA$\uparrow$ & IIR$\uparrow$ & Overall$\uparrow$ & VIA$\uparrow$ & IIR$\uparrow$ \\
\midrule
\multirow{6}{*}{Qwen3.5-122B-A10B} & Static & 63.16 & 55.39 & 68.13 & 28.33 & 35.00 & 22.33 & 84.38 & 50.00 & 50.00 \\
 & Restore & 65.11 & 42.11 & 82.24 & 5.75 & 14.00 & 1.00 & 95.05 & 50.00 & 50.00 \\
 & Reset & \textbf{72.40} & 38.88 & \textbf{99.93} & 15.83 & 12.50 & 3.00 & \textbf{96.38} & 0.00 & \textbf{100.00} \\
 & MemStrata & 66.10 & 55.53 & 73.02 & 28.66 & 32.00 & 22.50 & 84.43 & 50.00 & 50.00 \\
 & MemTX & 66.21 & \textbf{55.62} & 73.57 & 30.00 & 33.50 & 23.50 & 84.84 & 50.00 & 50.00 \\
 & TRACE & 65.20 & 53.60 & 72.36 & \textbf{52.33} & \textbf{52.50} & \textbf{50.50} & 94.32 & \textbf{95.96} & 98.78 \\
\midrule
\multirow{6}{*}{Gemini-3-flash} & Static & 67.63 & 54.46 & 76.08 & 70.31 & 76.27 & 63.77 & 88.04 & 50.00 & 50.00 \\
 & Restore & 66.87 & 45.00 & 83.43 & 7.77 & 19.21 & 0.52 & \textbf{98.40} & 50.00 & 50.00 \\
 & Reset & \textbf{72.00} & 43.31 & \textbf{95.23} & 8.83 & 20.49 & 0.52 & 97.75 & 0.00 & \textbf{100.00} \\
 & MemStrata & 67.83 & 55.53 & 75.74 & 64.80 & 74.88 & 56.71 & 86.79 & 50.00 & 50.00 \\
 & MemTX & 68.43 & \textbf{55.81} & 76.09 & 62.29 & 69.79 & 55.73 & 86.65 & 50.00 & 50.00 \\
 & TRACE & 67.70 & 54.41 & 76.41 & \textbf{88.81} & \textbf{89.99} & \textbf{87.73} & 98.32 & \textbf{98.34} & 99.47 \\
\midrule
\multirow{6}{*}{DeepSeek-v4-flash-0731} & Static & 68.36 & 52.47 & 80.20 & 22.05 & 34.20 & 12.62 & 65.45 & 50.00 & 50.00 \\
 & Restore & 69.81 & 42.27 & 91.13 & 6.02 & 16.03 & 0.00 & 93.51 & 50.00 & 50.00 \\
 & Reset & \textbf{72.51} & 39.97 & \textbf{99.27} & 16.09 & 23.55 & 0.00 & \textbf{96.39} & 0.00 & \textbf{100.00} \\
 & MemStrata & 69.31 & 53.57 & 80.01 & 26.00 & 36.23 & 14.58 & 76.18 & 50.00 & 50.00 \\
 & MemTX & 69.89 & \textbf{53.84} & 81.01 & 26.89 & 36.75 & 14.58 & 76.19 & 50.00 & 50.00 \\
 & TRACE & 68.83 & 51.86 & 80.72 & \textbf{54.40} & \textbf{57.81} & \textbf{49.94} & 92.63 & \textbf{92.58} & 98.44 \\
\bottomrule
\end{tabular*}
\end{table}

\subsection{Unified protocol and comparison methods}

Each episode has one returning agent and four active agents. All methods receive the same
post-absence checkpoint and workload in separate memory contexts, while failure manifests
preserve the eligible-episode denominator. The common harness compares \textsc{Static},
\textsc{Restore}, \textsc{Reset}, \textsc{MemStrata}, \textsc{MemTX}, and TRACE;
\textsc{MemStrata} and \textsc{MemTX} are mechanism-faithful reimplementations
\citep{yadav2026memstrata,li2026memtx}, and \textsc{CUPMem} is limited to the STALE
subset~\citep{chao2026stale}. Appendix~\ref{app:contract} gives settings and eligibility rules.
ManBench-Return balances Old-Valid and Old-Stale conditions after the active agents update the shared state.
Actors receive no gold labels, answer keys, relevant-session annotations, or judge rubrics during generation.

\subsection{Metrics}
\label{sec:metrics}

Overall is benchmark-specific accuracy. VIA and IIR measure valid-state availability
and invalid-state rejection, with WSAR as the complement of IIR. On STALE, VIA is IPA
accuracy, IIR jointly requires SR and PR, and LS requires all three probes in one episode
\citep{chao2026stale}; Memora uses native FAMA for retention and forgetting. Formal
definitions and aggregation weights appear in Appendices~\ref{app:metrics} and~\ref{app:memora-settings}.
Memora and ManBench use equal-weight condition averages, while Qwen STALE pools 64/64/72 episodes and Gemini/DeepSeek average strata equally.

\subsection{RQ1: memory validity and token cost under evolving context}
\label{sec:rq1}

\begin{table}[t]
\caption{Accuracy and token cost on the STALE Type~II subset.
Left: Overall (the mean of SR, PR, and IPA) and two native probe accuracies
(\%), with column maxima in bold.
Right: LS versus tokens per episode, with a magnified baseline cluster.
All methods are scored over the same 20 planned episodes per bucket; one
CUPMem episode exceeded the serving context window and is scored as a
failure on all three probes.}
\label{tab:stale-cost}
\centering
\begin{minipage}[t]{0.43\linewidth}
\vspace{0pt}
\centering
\small
\setlength{\tabcolsep}{2pt}
\renewcommand{\arraystretch}{1.45}
\begin{tabular*}{\linewidth}{@{\extracolsep{\fill}}lccc@{}}
\toprule
Method & Overall$\uparrow$ & PR$\uparrow$ & IPA$\uparrow$ \\
\midrule
Static & 33.33 & 26.67 & 45.00 \\
MemStrata & 31.11 & 36.67 & 35.00 \\
MemTX & 30.00 & 35.00 & 33.33 \\
CUPMem & \textbf{80.56} & \textbf{85.00} & \textbf{75.00} \\
TRACE (Ours) & 58.89 & 55.00 & 60.00 \\
\bottomrule
\end{tabular*}
\end{minipage}\hfill
\begin{minipage}[t]{0.54\linewidth}
\vspace{0pt}
\centering
\includegraphics[width=0.85\linewidth]{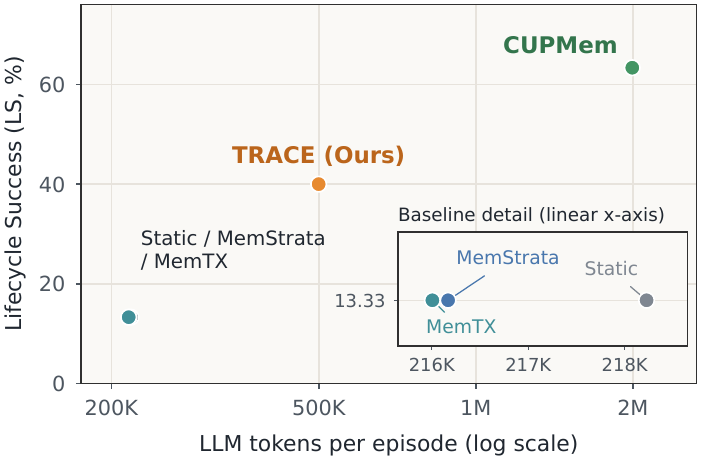}
\end{minipage}
\end{table}

Table~\ref{tab:main-results} reports actor-wise aggregates; condition-level grids and
aggregation details appear in Appendices~\ref{app:qwen}--\ref{app:deepseek}.

\paragraph{Memora: explicit replacement.}
TRACE does not consistently lead Memora. Averaged over the three departure
conditions, its FAMA is 40.14\%, 42.80\%, and 42.58\% with Qwen, Gemini,
and DeepSeek, respectively; \textsc{MemTX} obtains 42.14\%, 44.12\%, and 44.72\%.
At the 50\% departure point with Gemini, TRACE reaches the highest FAMA among the six methods
(43.51\%), but this advantage does not hold across all departure points.
The detailed grids in Appendices~\ref{app:qwen-memora},
\ref{app:gemini-memora}, and~\ref{app:deepseek-memora} therefore support a
limited conclusion: the return-time layer does not provide a uniform gain when explicit replacement
information is already available to the memory system.

\paragraph{STALE: implicit invalidation and token cost.}

In the six-method comparison, TRACE reaches 39.00\% LS with Qwen, 84.20\%
with Gemini, and 42.94\% with DeepSeek. The strongest comparison method for
each actor obtains 18.00\%, 59.26\%, and 11.63\%, respectively.
Since LS requires all three probes to pass in the same episode, these scores
measure joint state resolution and application. Available bucket results appear in
Appendices~\ref{app:qwen-stale}, \ref{app:gemini-stale}, and~\ref{app:deepseek-stale}.

We add a separate Qwen subset study to examine the accuracy--token trade-off
between return-time admission and \textsc{CUPMem}'s write-time state
consolidation~\citep{chao2026stale}, including the memory-maintenance cost
before answering. Table~\ref{tab:stale-cost} reports Overall alongside PR and
IPA. TRACE reaches 58.89\% Overall, 55.00\% PR, and 60.00\% IPA, exceeding
\textsc{Static}, \textsc{MemStrata}, and \textsc{MemTX} on all three reported scores. \textsc{CUPMem} obtains
higher scores, including 80.56\% Overall. Appendix~\ref{app:qwen-stale-subset}
also reports SR, the first component of Overall.

The right panel of Table~\ref{tab:stale-cost} relates LS to reported token use.
TRACE's 499,659 tokens
per episode are about 2.3 times the 216,006--218,227 used by the cheaper
methods; \textsc{CUPMem}'s 1,994,815 are 3.99 times TRACE's. These summaries place
TRACE between those methods and \textsc{CUPMem} in both accuracy and token use.
Appendix~\ref{app:qwen-stale-subset} gives the study rationale, bucket results,
and accounting details. Token counts do not measure wall-clock latency.

\begin{wrapfigure}{r}{0.46\textwidth}
\centering
\vspace{-6pt}
\includegraphics[width=\linewidth]{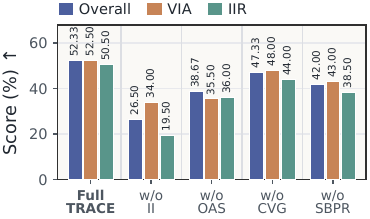}
\caption{Component ablations on STALE Type~II with Qwen (\%). Definitions and
full results: Appendix~\ref{app:core-ablation}.}
\label{fig:stale-core-ablation}
\vspace{-12pt}
\end{wrapfigure}
\paragraph{Component ablations.}
We ablate implicit invalidation (II), obligation-aware selection (OAS), the
complete-view gate (CVG), and source-bound private readmission (SBPR) in a
separate Qwen study on STALE Type~II (Figure~\ref{fig:stale-core-ablation}).
Full TRACE obtains 52.33\% Overall, 52.50\% VIA, and 50.50\% IIR. Removing II produces the
largest decline: Overall falls by 25.83 percentage points and IIR by 31.00
percentage points. Replacing OAS with relevance
ranking reduces Overall to 38.67\%; removing CVG instead
reduces it to 47.33\%. These interventions separate evidence selection from
the decision to proceed when coverage is incomplete. Removing SBPR
reduces Overall to 42.00\%, indicating that admission
also needs to constrain the private-memory retrieval path. Full TRACE has
the highest score on all three plotted metrics; Appendix~\ref{app:core-ablation}
reports the corresponding LS results.

\subsection{RQ2: state admission on ManBench-Return}

\textbf{ManBench-Return} exposes a gap between answer correctness and state admission.
\textsc{Reset} achieves 96.38\% Overall with Qwen and 96.39\% with DeepSeek, yet its VIA
is zero. TRACE attains lower Overall, 94.32\% with Qwen and 92.63\% with DeepSeek, while
preserving valid state: VIA and IIR are 95.96\% and 98.78\% with Qwen and 92.58\% and
98.44\% with DeepSeek, at WSAR of 1.22\% and 1.56\%. With Gemini, TRACE reaches 98.32\%
Overall against \textsc{Restore}'s 98.40\%, with 98.34\% VIA and 99.47\% IIR.
\textsc{Restore} attains only 50.00\% VIA and IIR in this aggregate, which illustrates
why answer correctness and state admission must be assessed separately.

\begin{figure}[t]
\centering
\vspace{-1.0em}
\includegraphics[width=\linewidth]{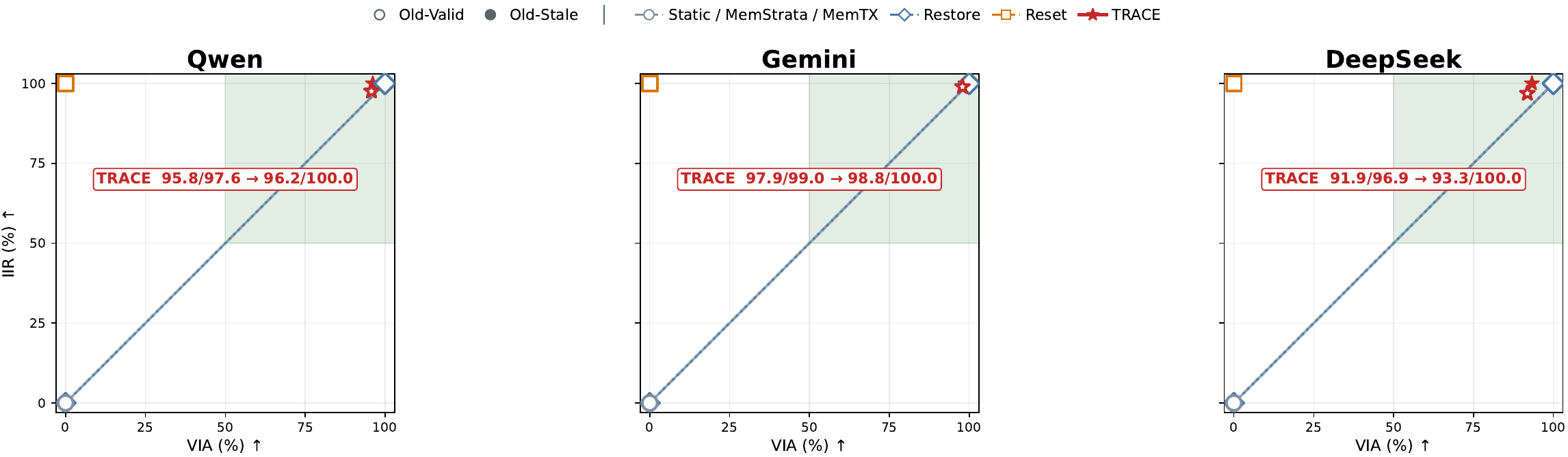}
\setlength{\abovecaptionskip}{3pt}
\caption{Admission plane for ManBench-Return. Markers show valid-state 
retention (VIA, horizontal) and invalid-state rejection (IIR, vertical) for 
each method and condition; upper-right is admissible. Hollow markers denote 
Old-Valid and filled markers Old-Stale. Reset conditions share the point
$(0,100)$, so their markers overlap.}
\label{fig:manbench-admission-plane}
\vspace{0.4em}
\end{figure}

The condition-level results show why. In Old-Valid, \textsc{Restore} reaches 100\% VIA
and IIR; in Old-Stale it reinstates invalid state and reaches 100\% WSAR.
\textsc{Static}, \textsc{MemStrata}, and \textsc{MemTX} show the reverse pattern, so
their 50\% aggregate VIA and IIR conceal opposite failures across conditions.
TRACE reaches 95.75\% VIA and 97.57\% IIR in Old-Valid and 96.17\% VIA and 100\% IIR in
Old-Stale; both preservation and rejection are needed to assess memory
admission (Figure~\ref{fig:manbench-admission-plane}).

While WSAR diagnoses context admission, ManBench's Reality Shift Rate (RSR) measures how
often an initially correct answer becomes incorrect after interaction
\citep{xu2026manbench}. In our return setting RSR applies to Old-Valid, where the task is
unchanged: Qwen's RSR is 29.94\% for \textsc{Static}, 6.63\% for TRACE, and 3.70\% for
\textsc{Restore}. A low WSAR therefore does not guarantee a correct final answer.
Appendix~\ref{app:manbench-settings} defines this adaptation; detailed condition grids for each actor appear in Appendices~\ref{app:qwen-manbench}, \ref{app:gemini-manbench}, and~\ref{app:deepseek-manbench}.

\subsection{RQ3: stability across team scale and memory backends}

\begin{table}[t]
\centering
\begin{minipage}[t]{0.47\linewidth}
\vspace{0pt}
\caption{ManBench-Return agent-scale diagnostic (\%). One agent returns while
the number of active agents varies.\strut}
\label{tab:agent-scale}
\centering
\small
\setlength{\tabcolsep}{3pt}
\renewcommand{\arraystretch}{1.22}
\begin{tabular*}{\linewidth}{@{\extracolsep{\fill}}clccc@{}}
\toprule
Agent & Method & Overall$\uparrow$ & VIA$\uparrow$ & IIR$\uparrow$ \\
\midrule
\multirow{2}{*}{4} & Static & 77.20 & 44.26 & 44.26 \\
& TRACE & \textbf{96.19} & \textbf{96.88} & \textbf{98.94} \\
\addlinespace[1pt]
\multirow{2}{*}{8} & Static & 76.61 & 44.20 & 44.20 \\
& TRACE & \textbf{96.04} & \textbf{96.63} & \textbf{98.72} \\
\addlinespace[1pt]
\multirow{2}{*}{16} & Static & 76.06 & 44.23 & 44.23 \\
& TRACE & \textbf{96.21} & \textbf{96.92} & \textbf{98.88} \\
\addlinespace[1pt]
\multirow{2}{*}{32} & Static & 76.18 & 44.26 & 44.26 \\
& TRACE & \textbf{95.85} & \textbf{96.62} & \textbf{98.76} \\
\bottomrule
\end{tabular*}
\end{minipage}\hfill
\begin{minipage}[t]{0.50\linewidth}
\vspace{0pt}
\centering
\captionof{figure}{Backend portability on Memora at 50\% departure (\%). Each
pair shares backend, actor, and scorer; VIA in
Table~\ref{tab:backend-portability}.\strut}
\label{fig:backend-portability}
\includegraphics[width=\linewidth]{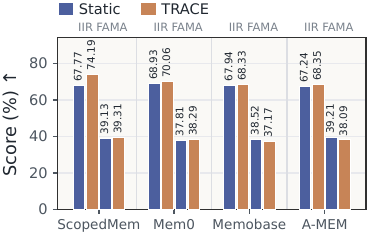}
\end{minipage}
\end{table}

\paragraph{Agent-scale diagnostic.} Varying the number of active agents over 4, 8,
16, and 32 while keeping one returning agent, TRACE's Overall, VIA, and IIR vary by
0.36, 0.30, and 0.22 percentage points (Table~\ref{tab:agent-scale}). The diagnostic supports
stability over this range (Appendix~\ref{app:agent-scale-study}); it does not establish increasing gains with team size.

\paragraph{Backend portability.} We attach TRACE to ScopedMem, Mem0, Memobase, and
A-MEM through a common memory-admission interface. Each backend supports a paired
\textsc{Static}/TRACE comparison on Memora at 50\% departure; we use Memora here because
portability concerns the integration boundary rather than the gain.
Figure~\ref{fig:backend-portability} reports IIR and FAMA for each paired
comparison.
IIR increases in all four pairs, by $+0.39$ to $+6.42$ points, while
VIA decreases by $0.05$ to $3.03$ points; FAMA changes range from $-1.35$ to
$+0.48$ percentage points. These results support interface portability with
mixed effects on answer quality.
Appendix~\ref{app:backend-study} describes the integration boundary, paired protocol,
and complete results in Table~\ref{tab:backend-portability}.

\section{Related Work}
\label{sec:related}

Our work connects three research directions: LLM-based multi-agent systems, persistent agent
memory, and temporal validity under changing context.
\paragraph{LLM-based multi-agent systems.}
Multi-agent systems coordinate through conversation, tool use, and shared
information~\citep{wu2024autogen,park2023generative}. ManBench shows how social
influence can reinforce incorrect beliefs~\citep{xu2026manbench}, while
AgentPoison and MAPLE-Guard study attacks on stored or propagated
memory~\citep{chen2024agentpoison,xiong2026mapleguard}. TRACE instead asks whether
previously valid state remains admissible when an agent returns to an evolved
workspace.

\paragraph{Persistent agent memory.}
External retrieval, memory tiers, and linked notes make historical information
available beyond the current context~\citep{lewis2020rag,packer2023memgpt,
xu2025amem,maharana2024locomo,wu2025longmemeval}. Memory construction and
retention use consolidation, admission, and query-conditioned
reinstatement~\citep{chhikara2025mem0,fang2026lightmem,zhang2026amac,yang2026ramem}.
TRACE studies the subsequent decision of whether an existing record may enter
a returning agent's context under the current task and state.

\paragraph{Temporal validity of memory.}
MemoryBank models forgetting through elapsed time and
importance~\citep{zhong2024memorybank}; Zep, \textsc{MemStrata}, and \textsc{MemTX} represent evolving
facts through temporal validity, commit checks, or dependent
repair~\citep{rasmussen2025zep,yadav2026memstrata,li2026memtx}. Memora and STALE
evaluate obsolete-memory reuse and implicit
invalidation~\citep{uddin2026memora,chao2026stale}. TRACE adds the return boundary,
combining invalidation with authorization, provenance, applicability, and
obligation coverage.

\section{Limitations and Future Work}

\paragraph{Limitations.} TRACE assumes an authenticated return, a preserved departure checkpoint, and source
receipts. Implicit invalidation depends on candidate coverage and verifier calibration:
missed dependencies admit stale state, over-broad checks discard useful context
(Appendix~\ref{app:counterexamples}). Performance degrades as absences lengthen: across the
STALE subset's Early and Late buckets, Overall falls from 73.33\% to 43.33\% while
tokens per episode rise from 316,750 to 664,860. ManBench-Return is derived, the agent-scale
study diagnostic, and the backend study tests portability rather than quality. The
token analysis reports run totals only, supporting no claim about latency, retry cost,
or peak context.

\paragraph{Future work.} We will test whether these effects persist across additional backends,
larger teams, longer return intervals, and open-ended tasks. We will study verifier calibration
under uncertainty or adversarial conditions while measuring
latency, retries, peak context, and token cost. Matched-budget verifier controls and
held-out return tasks will separate admission gains from extra verification information.
\emph{Proactive invalidation}---preemptively marking state stale after detected context shifts;
\emph{partial state recovery}---restoring invalidated components when dependencies are resolved; and
\emph{multi-agent memory reconciliation}---resolving conflicting absence-period updates through consensus.

\section{Conclusion}

Persistent storage preserves information, but it does not establish that stored state
remains safe after context changes. TRACE treats re-entry as a bounded admission event:
it reconciles the departure checkpoint with the current workspace, validates state,
covers obligations, and gates access to private memory. It shows no consistent
advantage under explicit updates, where direct contradictions make invalidation easy,
but improves the preservation--rejection balance over \textsc{Restore} and
\textsc{Reset} on STALE Type~II and ManBench-Return. These results support viewing
re-entry as a governance boundary with evidence, obligation coverage, and fail-closed
checks rather than an unconditional reload. Future work can test proactive invalidation,
partial recovery, and larger heterogeneous teams.

\subsection*{AI use statement}

Generative AI tools were used to assist with language editing, LaTeX drafting,
figure preparation, and implementation-oriented code inspection. The authors
checked the resulting text against the cited literature, verified numerical
claims against retained experiment artifacts, compiled and visually inspected
the manuscript, and remain responsible for all claims, code, and artifacts.

\subsection*{Ethics statement}

This work studies safeguards for persistent-memory agents and does not involve
human-subject experimentation. Return-time memory admission may reduce harmful
reuse of revoked or outdated information, but it can also suppress useful
context when provenance or dependency checks are incomplete. Deployments should
therefore expose rejection reasons, retain appropriate audit logs, protect
private memories, and provide a recovery path for authorized users rather than
treating automated admission decisions as infallible.

\subsection*{Reproducibility statement}

Section~\ref{sec:method} specifies the admission contract and its decision
stages; Section~\ref{sec:experiments} defines benchmark bindings, comparison
methods, and metrics. Appendix~\ref{app:contract} details the execution settings,
benchmark-specific return protocols, and aggregation rules; the remaining
appendices report detailed scores and evidence boundaries. The accompanying artifact preserves per-arm result
roots, frozen checkpoints, failure manifests, configuration receipts, and
reason-coded Return Views needed to audit the reported aggregates. An anonymous,
read-only mirror containing the implementation, frozen configurations, benchmark
inputs, and reproduction instructions is available at
\url{https://anonymous.4open.science/r/trace-review/}; Appendix~\ref{app:artifact}
summarizes its contents and entry points.

\bibliography{iclr2027_conference}

@inproceedings{lewis2020rag,
  title={Retrieval-Augmented Generation for Knowledge-Intensive NLP Tasks},
  author={Lewis, Patrick and Perez, Ethan and Piktus, Aleksandra and Petroni, Fabio and Karpukhin, Vladimir and Goyal, Naman and K{\"u}ttler, Heinrich and Lewis, Mike and Yih, Wen-tau and Rockt{\"a}schel, Tim and Riedel, Sebastian and Kiela, Douwe},
  booktitle={Advances in Neural Information Processing Systems},
  volume={33},
  pages={9459--9474},
  year={2020}
}

@inproceedings{wu2025longmemeval,
  title={{LongMemEval}: Benchmarking Chat Assistants on Long-Term Interactive Memory},
  author={Wu, Di and Wang, Hongwei and Yu, Wenhao and Zhang, Yuwei and Chang, Kai-Wei and Yu, Dong},
  booktitle={International Conference on Learning Representations},
  year={2025},
  url={https://arxiv.org/abs/2410.10813}
}

@inproceedings{xu2026manbench,
  title={When Agents ``Misremember'' Collectively: Exploring the Mandela Effect in LLM-based Multi-Agent Systems},
  author={Xu, Naen and An, Hengyu and Shi, Shuo and Zhang, Jinghuai and Zhou, Chunyi and Li, Changjiang and Du, Tianyu and Fu, Zhihui and Wang, Jun and Ji, Shouling},
  booktitle={International Conference on Learning Representations},
  year={2026},
  url={https://arxiv.org/abs/2602.00428}
}

@article{uddin2026memora,
  title={From Recall to Forgetting: Benchmarking Long-Term Memory for Personalized Agents},
  author={Uddin, Md Nayem and Shubham, Kumar and Blanco, Eduardo and Baral, Chitta and Wang, Gengyu},
  journal={arXiv preprint arXiv:2604.20006},
  year={2026}
}

@article{chao2026stale,
  title={{STALE}: Can LLM Agents Know When Their Memories Are No Longer Valid?},
  author={Chao, Hanxiang and Bai, Yihan and Sheng, Rui and Li, Tianle and Sun, Yushi},
  journal={arXiv preprint arXiv:2605.06527},
  year={2026}
}

@article{yadav2026memstrata,
  title={Temporal Validity in Retrieval Memory: Eliminating Stale-Fact Errors for AI Agents over Evolving Knowledge},
  author={Yadav, Neeraj},
  journal={arXiv preprint arXiv:2606.26511},
  year={2026}
}

@article{li2026memtx,
  title={{MemTX}: Transactional Belief Commit for Stateful Agent Memory},
  author={Li, Xiaoyang and Wang, Yiqi and Lu, Haohui and Chen, Zhi and Li, Mo and Song, Pingan and Zheng, Mingkai and Cai, Taotao},
  journal={arXiv preprint arXiv:2607.23929},
  year={2026}
}

@article{zhang2026amac,
  title={Adaptive Memory Admission Control for LLM Agents},
  author={Zhang, Guilin and Jiang, Wei and Wang, Xiejiashan and Behr, Aisha and Zhao, Kai and Friedman, Jeffrey and Chu, Xu and Anoun, Amine},
  journal={arXiv preprint arXiv:2603.04549},
  year={2026}
}

@article{yang2026ramem,
  title={{RaMem}: Contextual Reinstatement for Long-term Agentic Memory},
  author={Yang, Wei and Kan, Bryce and Li, Shixuan and Li, Li and Qin, Yuehan and Li, Jiate and Bogdan, Paul and Thomason, Jesse},
  journal={arXiv preprint arXiv:2606.22844},
  year={2026}
}

@inproceedings{park2023generative,
  title={Generative Agents: Interactive Simulacra of Human Behavior},
  author={Park, Joon Sung and O'Brien, Joseph C. and Cai, Carrie J. and Morris, Meredith Ringel and Liang, Percy and Bernstein, Michael S.},
  booktitle={Proceedings of the 36th Annual ACM Symposium on User Interface Software and Technology},
  publisher={Association for Computing Machinery},
  year={2023},
  url={https://arxiv.org/abs/2304.03442}
}

@inproceedings{shinn2023reflexion,
  title={Reflexion: Language Agents with Verbal Reinforcement Learning},
  author={Shinn, Noah and Cassano, Federico and Gopinath, Ashwin and Narasimhan, Karthik and Yao, Shunyu},
  booktitle={Advances in Neural Information Processing Systems},
  volume={36},
  year={2023},
  doi={10.52202/075280-0377}
}

@article{packer2023memgpt,
  title={{MemGPT}: Towards {LLMs} as Operating Systems},
  author={Packer, Charles and Wooders, Sarah and Lin, Kevin and Fang, Vivian and Patil, Shishir G. and Stoica, Ion and Gonzalez, Joseph E.},
  journal={arXiv preprint arXiv:2310.08560},
  year={2023},
  url={https://arxiv.org/abs/2310.08560}
}

@inproceedings{xu2025amem,
  title={{A-MEM}: Agentic Memory for {LLM} Agents},
  author={Xu, Wujiang and Liang, Zujie and Mei, Kai and Gao, Hang and Tan, Juntao and Zhang, Yongfeng},
  booktitle={Advances in Neural Information Processing Systems},
  year={2025},
  url={https://arxiv.org/abs/2502.12110}
}

@article{chhikara2025mem0,
  title={{Mem0}: Building Production-Ready {AI} Agents with Scalable Long-Term Memory},
  author={Chhikara, Prateek and Khant, Dev and Aryan, Saket and Singh, Taranjeet and Yadav, Deshraj},
  journal={arXiv preprint arXiv:2504.19413},
  year={2025},
  url={https://arxiv.org/abs/2504.19413}
}

@inproceedings{maharana2024locomo,
  title={Evaluating Very Long-Term Conversational Memory of {LLM} Agents},
  author={Maharana, Adyasha and Lee, Dong-Ho and Tulyakov, Sergey and Bansal, Mohit and Barbieri, Francesco and Fang, Yuwei},
  booktitle={Proceedings of the 62nd Annual Meeting of the Association for Computational Linguistics (Volume 1: Long Papers)},
  publisher={Association for Computational Linguistics},
  pages={13851--13870},
  year={2024},
  doi={10.18653/v1/2024.acl-long.747},
  url={https://aclanthology.org/2024.acl-long.747/}
}

@inproceedings{wu2024autogen,
  title={{AutoGen}: Enabling Next-Gen {LLM} Applications via Multi-Agent Conversation},
  author={Wu, Qingyun and Bansal, Gagan and Zhang, Jieyu and Wu, Yiran and Li, Beibin and Zhu, Erkang and Jiang, Li and Zhang, Xiaoyun and Zhang, Shaokun and Liu, Jiale and Awadallah, Ahmed Hassan and White, Ryen W. and Burger, Doug and Wang, Chi},
  booktitle={Conference on Language Modeling},
  year={2024},
  url={https://openreview.net/forum?id=BAakY1hNKS}
}

@article{rasmussen2025zep,
  title={{Zep}: A Temporal Knowledge Graph Architecture for Agent Memory},
  author={Rasmussen, Preston and Paliychuk, Pavlo and Beauvais, Travis and Ryan, Jack and Chalef, Daniel},
  journal={arXiv preprint arXiv:2501.13956},
  year={2025},
  url={https://arxiv.org/abs/2501.13956}
}

@inproceedings{zhong2024memorybank,
  title={{MemoryBank}: Enhancing Large Language Models with Long-Term Memory},
  author={Zhong, Wanjun and Guo, Lianghong and Gao, Qiqi and Ye, He and Wang, Yanlin},
  booktitle={Proceedings of the AAAI Conference on Artificial Intelligence},
  volume={38},
  pages={19724--19731},
  year={2024},
  doi={10.1609/aaai.v38i17.29946},
  url={https://ojs.aaai.org/index.php/AAAI/article/view/29946}
}

@article{liu2024lostmiddle,
  title={Lost in the Middle: How Language Models Use Long Contexts},
  author={Liu, Nelson F. and Lin, Kevin and Hewitt, John and Paranjape, Ashwin and Bevilacqua, Michele and Petroni, Fabio and Liang, Percy},
  journal={Transactions of the Association for Computational Linguistics},
  volume={12},
  pages={157--173},
  year={2024},
  doi={10.1162/tacl_a_00638},
  url={https://aclanthology.org/2024.tacl-1.9/}
}

@article{saltzer1975protection,
  title={The Protection of Information in Computer Systems},
  author={Saltzer, Jerome H. and Schroeder, Michael D.},
  journal={Proceedings of the IEEE},
  volume={63},
  number={9},
  pages={1278--1308},
  year={1975},
  doi={10.1109/PROC.1975.9939},
  url={https://web.mit.edu/Saltzer/www/publications/protection/}
}

@techreport{moreau2013prov,
  title={{PROV-DM}: The {PROV} Data Model},
  author={Moreau, Luc and Missier, Paolo},
  institution={World Wide Web Consortium},
  type={{W3C} Recommendation},
  year={2013},
  note={Editors. 30 April 2013},
  url={https://www.w3.org/TR/2013/REC-prov-dm-20130430/}
}

@article{chvatal1979greedy,
  title={A Greedy Heuristic for the Set-Covering Problem},
  author={Chv{\'a}tal, V.},
  journal={Mathematics of Operations Research},
  volume={4},
  number={3},
  pages={233--235},
  year={1979},
  doi={10.1287/moor.4.3.233},
  url={https://pubsonline.informs.org/doi/10.1287/moor.4.3.233}
}

@inproceedings{fang2026lightmem,
  title={{LightMem}: Lightweight and Efficient Memory-Augmented Generation},
  author={Fang, Jizhan and Deng, Xinle and Xu, Haoming and Jiang, Ziyan and Tang, Yuqi and Xu, Ziwen and Deng, Shumin and Yao, Yunzhi and Wang, Mengru and Qiao, Shuofei and Chen, Huajun and Zhang, Ningyu},
  booktitle={International Conference on Learning Representations},
  year={2026},
  url={https://arxiv.org/abs/2510.18866}
}

@article{doyle1979truth,
  title={A Truth Maintenance System},
  author={Doyle, Jon},
  journal={Artificial Intelligence},
  volume={12},
  number={3},
  pages={231--272},
  year={1979},
  doi={10.1016/0004-3702(79)90008-0},
  url={https://www.sciencedirect.com/science/article/pii/0004370279900080}
}

@misc{xiong2026mapleguard,
  title={{MAPLE-Guard}: Memory-Aware Link Enforcement Against Memory-Link Poisoning in Multi-Agent Systems},
  author={Xiong, Wenjun and Zhou, Yijin and Wang, Jiaqian and Gu, Shangding and Tang, Bo and Li, Zhiyu and Xiong, Feiyu and Wen, Ying and Wen, Muning},
  year={2026},
  eprint={2608.00426},
  archivePrefix={arXiv},
  primaryClass={cs.MA},
  url={https://arxiv.org/abs/2608.00426}
}

@inproceedings{chen2024agentpoison,
  title={{AgentPoison}: Red-teaming {LLM} Agents via Poisoning Memory or Knowledge Bases},
  author={Chen, Zhaorun and Xiang, Zhen and Xiao, Chaowei and Song, Dawn and Li, Bo},
  booktitle={Advances in Neural Information Processing Systems},
  volume={37},
  year={2024},
  doi={10.52202/079017-4136},
  url={https://arxiv.org/abs/2407.12784}
}

@techreport{hu2014abac,
  title={Guide to Attribute Based Access Control ({ABAC}) Definition and Considerations},
  author={Hu, Vincent and Ferraiolo, David and Kuhn, Richard and Schnitzer, Adam and Sandlin, Kenneth and Miller, Robert and Scarfone, Karen},
  institution={National Institute of Standards and Technology},
  type={NIST Special Publication},
  number={800-162},
  year={2014},
  doi={10.6028/NIST.SP.800-162},
  url={https://csrc.nist.gov/pubs/sp/800/162/upd2/final}
}

@misc{agentmemsurvey2024,
      title={A Survey on the Memory Mechanism of Large Language Model based Agents},
      author={Zeyu Zhang and Xiaohe Bo and Chen Ma and Rui Li and Xu Chen and Quanyu Dai and Jieming Zhu and Zhenhua Dong and Ji-Rong Wen},
      year={2024},
      eprint={2404.13501},
      archivePrefix={arXiv},
      primaryClass={cs.AI},
      url={https://arxiv.org/abs/2404.13501},
}

@misc{longmem2023,
      title={Augmenting Language Models with Long-Term Memory},
      author={Weizhi Wang and Li Dong and Hao Cheng and Xiaodong Liu and Xifeng Yan and Jianfeng Gao and Furu Wei},
      year={2023},
      eprint={2306.07174},
      archivePrefix={arXiv},
      primaryClass={cs.CL},
      url={https://arxiv.org/abs/2306.07174},
}

@misc{recurrentmem2023,
      title={Recurrent Memory Transformer},
      author={Aydar Bulatov and Yuri Kuratov and Mikhail S. Burtsev},
      year={2022},
      eprint={2207.06881},
      archivePrefix={arXiv},
      primaryClass={cs.CL},
      url={https://arxiv.org/abs/2207.06881},
}

@misc{agentbench2024,
      title={AgentBench: Evaluating LLMs as Agents},
      author={Xiao Liu and Hao Yu and Hanchen Zhang and Yifan Xu and Xuanyu Lei and Hanyu Lai and Yu Gu and Hangliang Ding and Kaiwen Men and Kejuan Yang and Shudan Zhang and Xiang Deng and Aohan Zeng and Zhengxiao Du and Chenhui Zhang and Sheng Shen and Tianjun Zhang and Yu Su and Huan Sun and Minlie Huang and Yuxiao Dong and Jie Tang},
      year={2024},
      eprint={2308.03688},
      archivePrefix={arXiv},
      primaryClass={cs.AI},
      url={https://arxiv.org/abs/2308.03688},
}

@misc{selfrag2024,
      title={Self-RAG: Learning to Retrieve, Generate, and Critique through Self-Reflection},
      author={Akari Asai and Zeqiu Wu and Yizhong Wang and Avirup Sil and Hannaneh Hajishirzi},
      year={2024},
      eprint={2310.11511},
      archivePrefix={arXiv},
      primaryClass={cs.CL},
      url={https://arxiv.org/abs/2310.11511},
}

@misc{crag2024,
      title={Corrective Retrieval Augmented Generation},
      author={Shi-Qi Yan and Jia-Chen Gu and Yun Zhu and Zhen-Hua Ling},
      year={2024},
      eprint={2401.15884},
      archivePrefix={arXiv},
      primaryClass={cs.CL},
      url={https://arxiv.org/abs/2401.15884},
}

@misc{raptor2024,
      title={RAPTOR: Recursive Abstractive Processing for Tree-Organized Retrieval},
      author={Parth Sarthi and Salman Abdullah and Aditi Tuli and Shubh Khanna and Anna Goldie and Christopher D. Manning},
      year={2024},
      eprint={2401.18059},
      archivePrefix={arXiv},
      primaryClass={cs.CL},
      url={https://arxiv.org/abs/2401.18059},
}

@misc{graphrag2024,
      title={From Local to Global: A Graph RAG Approach to Query-Focused Summarization},
      author={Darren Edge and Ha Trinh and Newman Cheng and Joshua Bradley and Alex Chao and Apurva Mody and Steven Truitt and Dasha Metropolitansky and Robert Osazuwa Ness and Jonathan Larson},
      year={2024},
      eprint={2404.16130},
      archivePrefix={arXiv},
      primaryClass={cs.CL},
      url={https://arxiv.org/abs/2404.16130},
}

@misc{hipporag2024,
      title={HippoRAG: Neurobiologically Inspired Long-Term Memory for Large Language Models},
      author={Bernal Jim{\'e}nez Guti{\'e}rrez and Yiheng Shu and Yu Gu and Michihiro Yasunaga and Yu Su},
      year={2024},
      eprint={2405.14831},
      archivePrefix={arXiv},
      primaryClass={cs.CL},
      url={https://arxiv.org/abs/2405.14831},
}

@misc{selfroute2024,
      title={Retrieval Augmented Generation or Long-Context LLMs? A Comprehensive Study and Hybrid Approach},
      author={Zhuowan Li and Cheng Li and Mingyang Zhang and Qiaozhu Mei and Michael Bendersky},
      year={2024},
      eprint={2407.16833},
      archivePrefix={arXiv},
      primaryClass={cs.CL},
      url={https://arxiv.org/abs/2407.16833},
}

@misc{gretriever2024,
      title={G-Retriever: Retrieval-Augmented Generation for Textual Graph Understanding and Question Answering},
      author={Xiaoxin He and Yijun Tian and Yifei Sun and Nitesh V. Chawla and Thomas Laurent and Yann LeCun and Xavier Bresson and Bryan Hooi},
      year={2024},
      eprint={2402.07630},
      archivePrefix={arXiv},
      primaryClass={cs.LG},
      url={https://arxiv.org/abs/2402.07630},
}

@misc{memoryllm2024,
      title={MEMORYLLM: Towards Self-Updatable Large Language Models},
      author={Yu Wang and Yifan Gao and Xiusi Chen and Haoming Jiang and Shiyang Li and Jingfeng Yang and Qingyu Yin and Zheng Li and Xian Li and Bing Yin and Jingbo Shang and Julian McAuley},
      year={2024},
      eprint={2402.04624},
      archivePrefix={arXiv},
      primaryClass={cs.CL},
      url={https://arxiv.org/abs/2402.04624},
}

@misc{memoryagentbench2026,
      title={Evaluating Memory in LLM Agents via Incremental Multi-Turn Interactions},
      author={Yuanzhe Hu and Yu Wang and Julian McAuley},
      year={2026},
      eprint={2507.05257},
      archivePrefix={arXiv},
      primaryClass={cs.CL},
      url={https://arxiv.org/abs/2507.05257},
}

@misc{rome2023,
      title={Locating and Editing Factual Associations in GPT},
      author={Kevin Meng and David Bau and Alex Andonian and Yonatan Belinkov},
      year={2022},
      eprint={2202.05262},
      archivePrefix={arXiv},
      primaryClass={cs.CL},
      url={https://arxiv.org/abs/2202.05262},
}

@misc{memit2023,
      title={Mass-Editing Memory in a Transformer},
      author={Kevin Meng and Arnab Sen Sharma and Alex Andonian and Yonatan Belinkov and David Bau},
      year={2023},
      eprint={2210.07229},
      archivePrefix={arXiv},
      primaryClass={cs.CL},
      url={https://arxiv.org/abs/2210.07229},
}

@misc{serac2022,
      title={Memory-Based Model Editing at Scale},
      author={Eric Mitchell and Charles Lin and Antoine Bosselut and Christopher D. Manning and Chelsea Finn},
      year={2022},
      eprint={2206.06520},
      archivePrefix={arXiv},
      primaryClass={cs.AI},
      url={https://arxiv.org/abs/2206.06520},
}

@misc{mquake2024,
      title={MQuAKE: Assessing Knowledge Editing in Language Models via Multi-Hop Questions},
      author={Zexuan Zhong and Zhengxuan Wu and Christopher D. Manning and Christopher Potts and Danqi Chen},
      year={2023},
      eprint={2305.14795},
      archivePrefix={arXiv},
      primaryClass={cs.CL},
      url={https://arxiv.org/abs/2305.14795},
}

@inproceedings{tan-etal-2025-membench,
    title = "{M}em{B}ench: Towards More Comprehensive Evaluation on the Memory of {LLM}-based Agents",
    author = "Tan, Haoran  and
      Zhang, Zeyu  and
      Ma, Chen  and
      Chen, Xu  and
      Dai, Quanyu  and
      Dong, Zhenhua",
    editor = "Che, Wanxiang  and
      Nabende, Joyce  and
      Shutova, Ekaterina  and
      Pilehvar, Mohammad Taher",
    booktitle = "Findings of the Association for Computational Linguistics: ACL 2025",
    month = jul,
    year = "2025",
    address = "Vienna, Austria",
    publisher = "Association for Computational Linguistics",
    url = "https://aclanthology.org/2025.findings-acl.989/",
    doi = "10.18653/v1/2025.findings-acl.989",
    pages = "19336--19352",
    ISBN = "979-8-89176-256-5"
}

@inproceedings{kang-etal-2025-memory,
    title = "Memory {OS} of {AI} Agent",
    author = "Kang, Jiazheng  and
      Ji, Mingming  and
      Zhao, Zhe  and
      Bai, Ting",
    editor = "Christodoulopoulos, Christos  and
      Chakraborty, Tanmoy  and
      Rose, Carolyn  and
      Peng, Violet",
    booktitle = "Proceedings of the 2025 Conference on Empirical Methods in Natural Language Processing",
    month = nov,
    year = "2025",
    address = "Suzhou, China",
    publisher = "Association for Computational Linguistics",
    url = "https://aclanthology.org/2025.emnlp-main.1318/",
    doi = "10.18653/v1/2025.emnlp-main.1318",
    pages = "25961--25970",
    ISBN = "979-8-89176-332-6"
}

@inproceedings{salama-etal-2025-meminsight,
    title = "{M}em{I}nsight: Autonomous Memory Augmentation for {LLM} Agents",
    author = "Salama, Rana  and
      Cai, Jason  and
      Yuan, Michelle  and
      Currey, Anna  and
      Sunkara, Monica  and
      Zhang, Yi  and
      Benajiba, Yassine",
    editor = "Christodoulopoulos, Christos  and
      Chakraborty, Tanmoy  and
      Rose, Carolyn  and
      Peng, Violet",
    booktitle = "Proceedings of the 2025 Conference on Empirical Methods in Natural Language Processing",
    month = nov,
    year = "2025",
    address = "Suzhou, China",
    publisher = "Association for Computational Linguistics",
    url = "https://aclanthology.org/2025.emnlp-main.1683/",
    doi = "10.18653/v1/2025.emnlp-main.1683",
    pages = "33136--33152",
    ISBN = "979-8-89176-332-6"
}

@inproceedings{tan-etal-2025-prospect,
    title = "In Prospect and Retrospect: Reflective Memory Management for Long-term Personalized Dialogue Agents",
    author = "Tan, Zhen  and
      Yan, Jun  and
      Hsu, I-Hung  and
      Han, Rujun  and
      Wang, Zifeng  and
      Le, Long  and
      Song, Yiwen  and
      Chen, Yanfei  and
      Palangi, Hamid  and
      Lee, George  and
      Iyer, Anand Rajan  and
      Chen, Tianlong  and
      Liu, Huan  and
      Lee, Chen-Yu  and
      Pfister, Tomas",
    editor = "Che, Wanxiang  and
      Nabende, Joyce  and
      Shutova, Ekaterina  and
      Pilehvar, Mohammad Taher",
    booktitle = "Proceedings of the 63rd Annual Meeting of the Association for Computational Linguistics (Volume 1: Long Papers)",
    month = jul,
    year = "2025",
    address = "Vienna, Austria",
    publisher = "Association for Computational Linguistics",
    url = "https://aclanthology.org/2025.acl-long.413/",
    doi = "10.18653/v1/2025.acl-long.413",
    pages = "8416--8439",
    ISBN = "979-8-89176-251-0"
}

@inproceedings{ong-etal-2025-towards,
    title = "Towards Lifelong Dialogue Agents via Timeline-based Memory Management",
    author = "Ong, Kai Tzu-iunn  and
      Kim, Namyoung  and
      Gwak, Minju  and
      Chae, Hyungjoo  and
      Kwon, Taeyoon  and
      Jo, Yohan  and
      Hwang, Seung-won  and
      Lee, Dongha  and
      Yeo, Jinyoung",
    editor = "Chiruzzo, Luis  and
      Ritter, Alan  and
      Wang, Lu",
    booktitle = "Proceedings of the 2025 Conference of the Nations of the Americas Chapter of the Association for Computational Linguistics: Human Language Technologies (Volume 1: Long Papers)",
    month = apr,
    year = "2025",
    address = "Albuquerque, New Mexico",
    publisher = "Association for Computational Linguistics",
    url = "https://aclanthology.org/2025.naacl-long.435/",
    doi = "10.18653/v1/2025.naacl-long.435",
    pages = "8631--8661",
    ISBN = "979-8-89176-189-6"
}

@inproceedings{jia-etal-2025-evaluating,
    title = "Evaluating the Long-Term Memory of Large Language Models",
    author = "Jia, Zixi  and
      Liu, Qinghua  and
      Li, Hexiao  and
      Chen, Yuyan  and
      Liu, Jiqiang",
    editor = "Che, Wanxiang  and
      Nabende, Joyce  and
      Shutova, Ekaterina  and
      Pilehvar, Mohammad Taher",
    booktitle = "Findings of the Association for Computational Linguistics: ACL 2025",
    month = jul,
    year = "2025",
    address = "Vienna, Austria",
    publisher = "Association for Computational Linguistics",
    url = "https://aclanthology.org/2025.findings-acl.1014/",
    doi = "10.18653/v1/2025.findings-acl.1014",
    pages = "19759--19777",
    ISBN = "979-8-89176-256-5"
}

@inproceedings{he-etal-2025-madial,
    title = "{MAD}ial-Bench: Towards Real-world Evaluation of Memory-Augmented Dialogue Generation",
    author = "He, Junqing  and
      Zhu, Liang  and
      Wang, Rui  and
      Wang, Xi  and
      Haffari, Gholamreza  and
      Zhang, Jiaxing",
    editor = "Chiruzzo, Luis  and
      Ritter, Alan  and
      Wang, Lu",
    booktitle = "Proceedings of the 2025 Conference of the Nations of the Americas Chapter of the Association for Computational Linguistics: Human Language Technologies (Volume 1: Long Papers)",
    month = apr,
    year = "2025",
    address = "Albuquerque, New Mexico",
    publisher = "Association for Computational Linguistics",
    url = "https://aclanthology.org/2025.naacl-long.499/",
    doi = "10.18653/v1/2025.naacl-long.499",
    pages = "9902--9921",
    ISBN = "979-8-89176-189-6"
}

@inproceedings{shen-etal-2026-mem2actbench,
    title = "{M}em2{A}ct{B}ench: A Benchmark for Evaluating Long-Term Memory Utilization in Task-Oriented Autonomous Agents",
    author = "Shen, Yiting  and
      Li, Kun  and
      Zhou, Wei  and
      Hu, Songlin",
    editor = "Liakata, Maria  and
      Moreira, Viviane P.  and
      Zhang, Jiajun  and
      Jurgens, David",
    booktitle = "Proceedings of the 64th Annual Meeting of the {A}ssociation for {C}omputational {L}inguistics (Volume 1: Long Papers)",
    month = jul,
    year = "2026",
    address = "San Diego, California, United States",
    publisher = "Association for Computational Linguistics",
    url = "https://aclanthology.org/2026.acl-long.370/",
    doi = "10.18653/v1/2026.acl-long.370",
    pages = "8173--8190",
    ISBN = "979-8-89176-390-6"
}

@inproceedings{du-etal-2024-perltqa,
    title = "{P}er{LTQA}: A Personal Long-Term Memory Dataset for Memory Classification, Retrieval, and Fusion in Question Answering",
    author = "Du, Yiming  and
      Wang, Hongru  and
      Zhao, Zhengyi  and
      Liang, Bin  and
      Wang, Baojun  and
      Zhong, Wanjun  and
      Wang, Zezhong  and
      Wong, Kam-Fai",
    editor = "Wong, Kam-Fai  and
      Zhang, Min  and
      Xu, Ruifeng  and
      Li, Jing  and
      Wei, Zhongyu  and
      Gui, Lin  and
      Liang, Bin  and
      Zhao, Runcong",
    booktitle = "Proceedings of the 10th SIGHAN Workshop on Chinese Language Processing (SIGHAN-10)",
    month = aug,
    year = "2024",
    address = "Bangkok, Thailand",
    publisher = "Association for Computational Linguistics",
    url = "https://aclanthology.org/2024.sighan-1.18/",
    pages = "152--164"
}

@inproceedings{bei-etal-2026-mem,
    title = "Mem-Gallery: Benchmarking Multimodal Long-Term Conversational Memory for {MLLM} Agents",
    author = "Bei, Yuanchen  and
      Wei, Tianxin  and
      Ning, Xuying  and
      Zhao, Yanjun  and
      Liu, Zhining  and
      Lin, Xiao  and
      Zhu, Yada  and
      Hamann, Hendrik  and
      He, Jingrui  and
      Tong, Hanghang",
    editor = "Liakata, Maria  and
      Moreira, Viviane P.  and
      Zhang, Jiajun  and
      Jurgens, David",
    booktitle = "Proceedings of the 64th Annual Meeting of the {A}ssociation for {C}omputational {L}inguistics (Volume 1: Long Papers)",
    month = jul,
    year = "2026",
    address = "San Diego, California, United States",
    publisher = "Association for Computational Linguistics",
    url = "https://aclanthology.org/2026.acl-long.1892/",
    doi = "10.18653/v1/2026.acl-long.1892",
    pages = "40750--40784",
    ISBN = "979-8-89176-390-6"
}

@inproceedings{banerjee-etal-2026-apex,
    title = "{APEX}-{MEM}: Agentic Semi-Structured Memory with Temporal Reasoning for Long-Term Conversational {AI}",
    author = "Banerjee, Pratyay  and
      Moshtaghi, Masud  and
      Subramanian, Shivashankar  and
      Misra, Amita  and
      Chadha, Ankit",
    editor = "Liakata, Maria  and
      Moreira, Viviane P.  and
      Zhang, Jiajun  and
      Jurgens, David",
    booktitle = "Proceedings of the 64th Annual Meeting of the {A}ssociation for {C}omputational {L}inguistics (Volume 1: Long Papers)",
    month = jul,
    year = "2026",
    address = "San Diego, California, United States",
    publisher = "Association for Computational Linguistics",
    url = "https://aclanthology.org/2026.acl-long.749/",
    doi = "10.18653/v1/2026.acl-long.749",
    pages = "16470--16489",
    ISBN = "979-8-89176-390-6"
}

@inproceedings{li-etal-2026-timem,
    title = "{T}i{M}em: Temporal-Hierarchical Memory Consolidation for Long-Horizon Conversational Agents",
    author = "Li, Kai  and
      Yu, Xuanqing  and
      Ni, Ziyi  and
      Zeng, Yi  and
      Xu, Yao  and
      Zhang, Zheqing  and
      Li, Xin  and
      Sang, Jitao  and
      Duan, Xiaogang  and
      Wang, Xuelei  and
      Liu, Chengbao  and
      Tan, Jie",
    editor = "Liakata, Maria  and
      Moreira, Viviane P.  and
      Zhang, Jiajun  and
      Jurgens, David",
    booktitle = "Findings of the {A}ssociation for {C}omputational {L}inguistics: {ACL} 2026",
    month = jul,
    year = "2026",
    address = "San Diego, California, United States",
    publisher = "Association for Computational Linguistics",
    url = "https://aclanthology.org/2026.findings-acl.1091/",
    doi = "10.18653/v1/2026.findings-acl.1091",
    pages = "21700--21720",
    ISBN = "979-8-89176-395-1"
}

@inproceedings{ye-etal-2026-h,
    title = "{H}-Mem: Hybrid Multi-Dimensional Memory Management for Long-Context Conversational Agents",
    author = "Ye, Zihe  and
      Huang, Jingyuan  and
      Chen, Weixin  and
      Zhang, Yongfeng",
    editor = "Demberg, Vera  and
      Inui, Kentaro  and
      Marquez, Llu{\'i}s",
    booktitle = "Proceedings of the 19th Conference of the {E}uropean Chapter of the {A}ssociation for {C}omputational {L}inguistics (Volume 1: Long Papers)",
    month = mar,
    year = "2026",
    address = "Rabat, Morocco",
    publisher = "Association for Computational Linguistics",
    url = "https://aclanthology.org/2026.eacl-long.363/",
    doi = "10.18653/v1/2026.eacl-long.363",
    pages = "7756--7775",
    ISBN = "979-8-89176-380-7"
}

@inproceedings{latimer-etal-2026-hindsight,
    title = "Hindsight: Structured Agent Memory that Retains, Recalls, and Reflects",
    author = "Latimer, Christopher  and
      Boschi, Nicol{\`o}  and
      Neeser, Andrew  and
      Bartholomew, Chris  and
      Srivastava, Gaurav  and
      Wang, Xuan  and
      Ramakrishnan, Naren",
    editor = "Durrett, Greg  and
      Jian, Ping",
    booktitle = "Proceedings of the 64th Annual Meeting of the {A}ssociation for {C}omputational {L}inguistics (Volume 3: System Demonstrations)",
    month = jul,
    year = "2026",
    address = "San Diego, California, United States",
    publisher = "Association for Computational Linguistics",
    url = "https://aclanthology.org/2026.acl-demo.27/",
    doi = "10.18653/v1/2026.acl-demo.27",
    pages = "275--285",
    ISBN = "979-8-89176-392-0"
}

@inproceedings{jiang-etal-2023-active,
    title = "Active Retrieval Augmented Generation",
    author = "Jiang, Zhengbao  and
      Xu, Frank  and
      Gao, Luyu  and
      Sun, Zhiqing  and
      Liu, Qian  and
      Dwivedi-Yu, Jane  and
      Yang, Yiming  and
      Callan, Jamie  and
      Neubig, Graham",
    editor = "Bouamor, Houda  and
      Pino, Juan  and
      Bali, Kalika",
    booktitle = "Proceedings of the 2023 Conference on Empirical Methods in Natural Language Processing",
    month = dec,
    year = "2023",
    address = "Singapore",
    publisher = "Association for Computational Linguistics",
    url = "https://aclanthology.org/2023.emnlp-main.495/",
    doi = "10.18653/v1/2023.emnlp-main.495",
    pages = "7969--7992"
}

@inproceedings{dhuliawala-etal-2024-chain,
    title = "Chain-of-Verification Reduces Hallucination in Large Language Models",
    author = "Dhuliawala, Shehzaad  and
      Komeili, Mojtaba  and
      Xu, Jing  and
      Raileanu, Roberta  and
      Li, Xian  and
      Celikyilmaz, Asli  and
      Weston, Jason",
    editor = "Ku, Lun-Wei  and
      Martins, Andre  and
      Srikumar, Vivek",
    booktitle = "Findings of the Association for Computational Linguistics: ACL 2024",
    month = aug,
    year = "2024",
    address = "Bangkok, Thailand",
    publisher = "Association for Computational Linguistics",
    url = "https://aclanthology.org/2024.findings-acl.212/",
    doi = "10.18653/v1/2024.findings-acl.212",
    pages = "3563--3578"
}

@inproceedings{gao-etal-2023-rarr,
    title = "{RARR}: Researching and Revising What Language Models Say, Using Language Models",
    author = "Gao, Luyu  and
      Dai, Zhuyun  and
      Pasupat, Panupong  and
      Chen, Anthony  and
      Chaganty, Arun Tejasvi  and
      Fan, Yicheng  and
      Zhao, Vincent  and
      Lao, Ni  and
      Lee, Hongrae  and
      Juan, Da-Cheng  and
      Guu, Kelvin",
    editor = "Rogers, Anna  and
      Boyd-Graber, Jordan  and
      Okazaki, Naoaki",
    booktitle = "Proceedings of the 61st Annual Meeting of the Association for Computational Linguistics (Volume 1: Long Papers)",
    month = jul,
    year = "2023",
    address = "Toronto, Canada",
    publisher = "Association for Computational Linguistics",
    url = "https://aclanthology.org/2023.acl-long.910/",
    doi = "10.18653/v1/2023.acl-long.910",
    pages = "16477--16508"
}

@inproceedings{min-etal-2023-factscore,
    title = "{FA}ct{S}core: Fine-grained Atomic Evaluation of Factual Precision in Long Form Text Generation",
    author = "Min, Sewon  and
      Krishna, Kalpesh  and
      Lyu, Xinxi  and
      Lewis, Mike  and
      Yih, Wen-tau  and
      Koh, Pang  and
      Iyyer, Mohit  and
      Zettlemoyer, Luke  and
      Hajishirzi, Hannaneh",
    editor = "Bouamor, Houda  and
      Pino, Juan  and
      Bali, Kalika",
    booktitle = "Proceedings of the 2023 Conference on Empirical Methods in Natural Language Processing",
    month = dec,
    year = "2023",
    address = "Singapore",
    publisher = "Association for Computational Linguistics",
    url = "https://aclanthology.org/2023.emnlp-main.741/",
    doi = "10.18653/v1/2023.emnlp-main.741",
    pages = "12076--12100"
}

@inproceedings{manakul-etal-2023-selfcheckgpt,
    title = "{S}elf{C}heck{GPT}: Zero-Resource Black-Box Hallucination Detection for Generative Large Language Models",
    author = "Manakul, Potsawee  and
      Liusie, Adian  and
      Gales, Mark",
    editor = "Bouamor, Houda  and
      Pino, Juan  and
      Bali, Kalika",
    booktitle = "Proceedings of the 2023 Conference on Empirical Methods in Natural Language Processing",
    month = dec,
    year = "2023",
    address = "Singapore",
    publisher = "Association for Computational Linguistics",
    url = "https://aclanthology.org/2023.emnlp-main.557/",
    doi = "10.18653/v1/2023.emnlp-main.557",
    pages = "9004--9017"
}

@article{alchourron1985logic,
  title={On the Logic of Theory Change: Partial Meet Contraction and Revision Functions},
  author={Alchourr{\'o}n, Carlos E. and G{\"a}rdenfors, Peter and Makinson, David},
  journal={The Journal of Symbolic Logic},
  volume={50},
  number={2},
  pages={510--530},
  year={1985},
  doi={10.2307/2274239}
}

@article{reiter1980default,
  title={A Logic for Default Reasoning},
  author={Reiter, Raymond},
  journal={Artificial Intelligence},
  volume={13},
  number={1--2},
  pages={81--132},
  year={1980},
  doi={10.1016/0004-3702(80)90014-4}
}

@article{dekleer1986assumption,
  title={An Assumption-Based TMS},
  author={de Kleer, Johan},
  journal={Artificial Intelligence},
  volume={28},
  number={2},
  pages={127--162},
  year={1986},
  doi={10.1016/0004-3702(86)90080-9}
}

@article{simmhan2005survey,
  title={A Survey of Data Provenance in E-Science},
  author={Simmhan, Yogavaram and Plale, Beth and Gannon, Dennis},
  journal={SIGMOD Record},
  volume={34},
  number={3},
  pages={31--36},
  year={2005},
  doi={10.1145/1084805.1084812}
}

@inproceedings{buneman2001provenance,
  title={Why and Where: A Characterization of Data Provenance},
  author={Buneman, Peter and Khanna, Sanjeev and Tan, Wang-Chiew},
  booktitle={International Conference on Database Theory},
  pages={316--330},
  year={2001},
  publisher={Springer},
  doi={10.1007/3-540-44503-X_20}
}

@book{snodgrass1999temporal,
  title={Developing Time-Oriented Database Applications in {SQL}},
  author={Snodgrass, Richard T.},
  publisher={Morgan Kaufmann},
  year={1999},
  address={San Francisco, CA}
}

@article{sandhu1996rbac,
  title={Role-Based Access Control Models},
  author={Sandhu, Ravi S. and Coyne, Edward J. and Feinstein, Hal L. and Youman, Charles E.},
  journal={IEEE Computer},
  volume={29},
  number={2},
  pages={38--47},
  year={1996},
  doi={10.1109/2.485845}
}
\bibliographystyle{iclr2027_conference}

\clearpage
\appendix

\section{Reproducibility Artifact}
\label{app:artifact}

The anonymous, read-only artifact is available at
\url{https://anonymous.4open.science/r/trace-review/}. It contains the
TRACE implementation, frozen configurations, maintained benchmark inputs, vendored
notices, and reproduction instructions. API credentials, raw provider logs, and
author repository history are excluded; the artifact starts from the offline
regression command in \texttt{REPRODUCIBILITY.md} and the maintained entry points
listed in \texttt{scripts/experiments/README.md}.

\section{Related Work}
\label{app:related-work}

\subsection{Memory architectures and continual benchmarks.}
Surveys and cognitive architectures distinguish working, episodic, semantic, and
procedural memory, and describe how an agent can write, retrieve, and reflect over
external state~\citep{agentmemsurvey2024,longmem2023,recurrentmem2023,memoryllm2024}.
Experience-oriented systems reuse trajectories or
skills~\citep{packer2023memgpt,xu2025amem}, while newer
managers organize, summarize, or time-index long-term conversations
~\citep{kang-etal-2025-memory,salama-etal-2025-meminsight,
tan-etal-2025-prospect,ong-etal-2025-towards,banerjee-etal-2026-apex,
li-etal-2026-timem,ye-etal-2026-h,latimer-etal-2026-hindsight}. These systems
improve what can be stored or recalled; they do not by themselves define the
return-time authority of a record after the workspace has changed.

LoCoMo and LongMemEval measure long-range conversational recall
~\citep{maharana2024locomo,wu2025longmemeval}, while Memora and STALE test
forgetting, conflicting updates, and implicit invalidation~\citep{uddin2026memora,
chao2026stale}. More recent benchmarks separate factual memory, task use, and
continual repair~\citep{tan-etal-2025-membench,memoryagentbench2026,
shen-etal-2026-mem2actbench,du-etal-2024-perltqa,he-etal-2025-madial,
jia-etal-2025-evaluating,bei-etal-2026-mem}. They expose the same broad need for
selective retention and conflict resolution, but do not impose TRACE's authenticated
return contract or its joint validity-and-obligation metrics.

\subsection{Multi-agent coordination and evaluation.}
LLM teams coordinate through role prompts, conversational protocols, tool use, and
structured task graphs~\citep{wu2024autogen}. AgentBench evaluates such systems
in interactive environments with tools, users, or adversarial inputs
~\citep{agentbench2024}. These studies establish the importance of
long-horizon coordination and failure analysis; TRACE focuses on the narrower
boundary at which one participant rejoins and must decide which shared state is
still admissible.

\subsection{Retrieval, evidence, and model updates.}
Retrieval-augmented systems range from parametric retrievers to adaptive retrieval
and long-context routing~\citep{lewis2020rag,liu2024lostmiddle,jiang-etal-2023-active,selfroute2024}. Self-RAG and CRAG add retrieval
critique or correction~\citep{selfrag2024,crag2024}, while RAPTOR, GraphRAG,
HippoRAG, and G-Retriever organize evidence hierarchically or as graphs
~\citep{raptor2024,graphrag2024,hipporag2024,gretriever2024}. These methods improve
candidate recall and context efficiency. TRACE applies an additional admission
decision after retrieval, so a relevant candidate can still be rejected when its
authorization, temporal validity, provenance, or task applicability fails.

FActScore, SelfCheckGPT, RARR, and Chain-of-Verification assess atomic support,
consistency, attribution, or post-hoc verification~\citep{min-etal-2023-factscore,
manakul-etal-2023-selfcheckgpt,gao-etal-2023-rarr,dhuliawala-etal-2024-chain}.
Parametric editing methods such as ROME, MEMIT, and SERAC change or cache model
associations, and MQuAKE tests whether such edits propagate through multi-hop
consequences~\citep{rome2023,memit2023,serac2022,mquake2024}. These approaches
provide useful evidence and update primitives, but they are not return-time
authorization policies and do not replace the external memory backend.

\subsection{Governance foundations.}
Attribute- and role-based access control provide the authorization vocabulary
~\citep{hu2014abac,sandhu1996rbac}; PROV and provenance surveys describe how to
bind a claim to its source~\citep{moreau2013prov,simmhan2005survey,buneman2001provenance}.
Temporal database models make validity intervals explicit~\citep{snodgrass1999temporal},
while AGM belief revision, default logic, assumption-based truth maintenance, and
classical truth-maintenance systems formalize revision under changing evidence
~\citep{alchourron1985logic,reiter1980default,dekleer1986assumption,doyle1979truth}.
Fail-closed protection principles motivate rejecting an item when a required check
cannot be established~\citep{saltzer1975protection}. TRACE combines these foundations
at a single lifecycle boundary and adds obligation coverage and a bounded Return View
to the existing storage and retrieval pipeline.

\section{Method Details and Case Studies}
\label{app:method-details}
\label{app:evidence}

This section elaborates the admission procedure in Section~\ref{sec:method}.
It distinguishes source provenance from current validity, defines obligation
coverage, and explains how the selected work state controls private-memory
reuse. Section~\ref{app:return-case} follows a complete return decision in
the travel example from Figure~\ref{fig:return-dilemma}.
Sections~\ref{app:benchmark-case} and~\ref{app:counterexamples} provide
observed benchmark cases, including failures of invalidation and subsequent
use of admitted evidence.

\subsection{Memory records, provenance, and current applicability}

TRACE operates on work-state statements supplied through a memory adapter.
Each statement has a task and role scope, a validity interval, and a binding
to its source evidence. Three relations have distinct meanings: a provenance
relation identifies where a statement came from; an invalidation relation
identifies a later event that changes its current applicability; and a support
relation identifies the evidence requirement served by the statement. A
source binding therefore cannot substitute for a validity decision. An
authentic booking confirmation, for example, remains traceable after the
booking has been cancelled.

The predicates in Equation~\ref{eq:eligible} address separate questions.
Authorization determines whether the returning principal and role may access
the item. Temporal validity checks whether its applicability extends to the
return state. Task applicability checks the current task, purpose, and item
type. Provenance binds the item and its claimed support relations to source
evidence. This separation follows the distinction between attribute-based
access policies~\citep{hu2014abac} and provenance
descriptions~\citep{moreau2013prov}. A provenance check establishes a source association;
it does not independently prove that the source statement is true.

Private episodic memories remain in the owning agent's bank. The shared
admission procedure selects work-state items, and the resulting source set
determines which private experiences may subsequently be retrieved. This
prevents an experience rejected at the work-state level from re-entering
through a separate private-memory retrieval path.

\subsection{From a possible dependency to a confirmed invalidation}

Explicit revocation and supersession relations can be applied directly.
Implicit invalidation requires an additional inference: whether a later
observation changes a premise needed to use an earlier statement in the
current state. The scanner proposes such pairs broadly. Its output is a
set of hypotheses, not a decision to discard every related memory.

This distinction is related to truth-maintenance systems, which revise beliefs
when their justifications change~\citep{doyle1979truth}. TRACE applies the same
idea to source-bound, task-conditioned admission at an agent's return.

The verifier then examines each proposed pair using the original observations
and their temporal order. It asks whether treating the earlier statement as
current would be incompatible with the later evidence. Merely sharing a
topic is insufficient. For example, a later discussion of train travel does
not invalidate an existing itinerary, whereas a confirmed change of
destination can invalidate a recommendation tied to the former destination.
The earlier recommendation may remain historically correct in either case.

In the STALE adapter, scanning and verification precede the downstream
questions and use only the supplied observation pairs, without benchmark
answers or conflict annotations. A confirmed invalidation excludes the older
item from current use; it does not automatically authorize the later item,
which still undergoes the item-level checks. The scan and verifier are model
judgments, so this design does not guarantee that every implicit dependency
will be found. Separating the stages allows broad candidate generation
without equating a possible dependency with established invalidation.

\subsection{Obligation coverage and budgeted selection}

An obligation specifies information needed for the returning role to continue
its task. It is accompanied by a progress frontier describing what remains
unresolved. The task adapter supplies these requirements and the relations
linking them to candidate evidence. Coverage measures whether that evidence
is available in the Return View; it does not mean that the agent has already
completed the obligation.

For each obligation $o\in\mathcal O$, let $R_o$ be its required dependency
set and $K_o\subseteq R_o$ its critical dependencies. If $D(m)$ denotes the
dependencies supported by item $m$, the evidence available from a selected
set $S$ is $D(S)=\bigcup_{m\in S}D(m)$. The coverage measures are
\begin{equation}
 \operatorname{Cov}(S)=
 \frac{1}{|\mathcal O|}\sum_{o\in\mathcal O}
 \mathbf{1}\!\left[R_o\subseteq D(S)\right],
 \qquad
 \operatorname{Crit}(S)=\frac{|D(S)\cap K|}{|K|},
 \quad K=\bigcup_{o\in\mathcal O}K_o.
 \label{eq:coverage}
\end{equation}
An obligation counts as covered only when all of its required dependencies
are present. Critical coverage counts distinct critical dependencies, even
when several obligations share them. An empty obligation or critical set
has coverage one by convention.

The selector first reserves space for the declared open obligations and
their progress frontiers. It then adds eligible evidence according to its
newly covered dependencies per unit of disclosure cost:
\begin{equation}
 g(m\mid S)=
 \frac{\displaystyle\sum_{d\in(D(m)\cap R)\setminus D(S)}w(d)}{c(m)},
 \qquad R=\bigcup_{o\in\mathcal O}R_o,
 \qquad
 w(d)=\begin{cases}2,&d\in K,\\1,&d\notin K.\end{cases}
 \label{eq:selection-gain}
\end{equation}
Here $c(m)=\max\{1,\ell(m)\}$, with $\ell(m)$ the item's character length.
This cost controls Return View size; it differs from the total LLM token
consumption reported in the experiments.
Each addition must respect both item-count and character budgets. The
selector prefers greater uncovered support, with extra weight on critical
dependencies, rather than longer or more recent records. This is a
constrained greedy heuristic related to weighted set
cover~\citep{chvatal1979greedy}; the hard budgets and mandatory structure preclude
a direct use of the classical approximation guarantee.

If the mandatory structure is absent or the selected set fails the required
coverage thresholds, TRACE does not issue that set as a sufficient Return
View. Thus individually admissible items can still form an inadequate
context, and a high retrieval score cannot compensate for missing critical
evidence.

\subsection{Return View validation and incomplete admission}

A Return View binds the selected content to one return event, principal,
role, task, and state epoch. Authentication protects this binding and allows
the admission layer to detect modification or use in an incompatible return
context. It does not certify the factual truth of the content. Before
exposure, TRACE rechecks the binding, item eligibility, coverage, and context
budgets against the current return conditions.

When admission is incomplete, the configured policy either resets the
returning context or blocks the return. Reset supplies no inherited memory;
blocking withholds admission until sufficient context is available. Neither
outcome treats a partial view as obligation-complete. This policy follows
the fail-safe-default principle~\citep{saltzer1975protection}.
Once a view is admitted, a private experience is eligible for readmission
only if its bound work-state source belongs to that view. Excluded content
may remain stored, but it is unavailable as inherited context. This is
functional forgetting at the retrieval boundary, without modifying model
parameters or requiring physical deletion from the backend.

\subsection{Scope of the method}

TRACE assumes trustworthy lifecycle identity and source bindings. Its
evaluation addresses explicit updates, implicit invalidation, and propagated
wrong state; it does not establish resistance to arbitrary memory poisoning
or forged evidence. Memory-poisoning defenses such as MAPLE-Guard instead
intervene across writing, retrieval, promotion, and cross-agent
reuse~\citep{xiong2026mapleguard}. These mechanisms concern a broader adversarial
memory lifecycle, while TRACE evaluates admission at an agent's return.
The scale study supports conclusions over the tested team sizes, and the
backend study establishes interface portability across four stores rather
than uniform accuracy gains. The MemStrata and MemTX comparisons use
mechanism-faithful adaptations within the common evaluation harness.

\newtcolorbox{tracecasepanel}[1]{
  colback=black!4,
  colframe=black!60,
  colbacktitle=black!60,
  coltitle=white,
  fonttitle=\bfseries,
  title={#1},
  boxrule=0.5pt,
  arc=3mm,
  left=9pt,right=9pt,top=7pt,bottom=7pt,
  toptitle=4pt,bottomtitle=4pt,
  before skip=10pt,after skip=0pt,
  before upper={\setlength{\parindent}{0pt}\setlength{\parskip}{5pt}}
}
\newcommand{\tracecaseoutcome}{\par\nobreak\vspace{5pt}{\color{black!60}\hrule height 0.4pt}\vspace{4pt}}

\subsection{Case study: returning to a changed travel plan}
\label{app:return-case}

This illustrative case follows Figure~\ref{fig:return-dilemma} from departure
to an admitted Return View. A1 coordinates the itinerary, A2 manages bookings,
A3 checks the budget, and A4 arranges transport. The final pickup confirmation
is a hypothetical continuation; the case explains the mechanism and is not
an additional benchmark result.

\begin{tracecasepanel}{1. Before departure: a coherent stored plan}
\textbf{Family requirements ($c$).}
A quiet hotel near the station, with a total hotel cost below \$300.

\textbf{A1's stored itinerary.}
Sunrise Hotel is booked ($h_0$); the train departs at 9:00 ($t_0$); the
recommended pickup is at 8:00 from Sunrise ($p_0$).

\textbf{Departure boundary.}
The checkpoint preserves these records, their source bindings, and A1's
identity, role, and task scope. A1's private experience remains in its own
bank and is unavailable to the fresh workload until readmission.
\tracecaseoutcome
\textbf{State at departure:} the itinerary is coherent, and the family
requirements remain relevant to the trip.
\end{tracecasepanel}

\begin{tracecasepanel}{2. During absence: the team changes the plan}
\textbf{A2 / booking update.}
Sunrise cancels the reservation. A2 confirms Riverside as its replacement
for \$282 ($h_1$).

\textbf{A4 / schedule update.}
The train now departs at 9:30 ($t_1$). The team must still confirm a new
pickup location and time.

\textbf{Returning task.}
A1 rejoins under the same identity and itinerary-planning role, authorized
to read the shared trip records and prepare the final hotel and transport
details for the family.
\tracecaseoutcome
\textbf{Candidate context:} departure records plus the absence-period
updates, including the unresolved pickup obligation and its progress frontier.
\end{tracecasepanel}

\begin{tracecasepanel}{3. At return: determine which records remain admissible}
\textbf{Explicit invalidation.}
The cancellation and replacement exclude $h_0$; the revised train schedule
supersedes $t_0$.

\textbf{Implicit invalidation.}
The scanner links $p_0$ to the hotel change. The verifier confirms that its
Sunrise pickup location conflicts with the Riverside booking. A later train
time alone would not make an early pickup invalid; the changed location is
the decisive evidence.

\textbf{Surviving evidence.}
The family constraints $c$ are unchanged. The replacement booking $h_1$ and
schedule $t_1$ pass the authorization, temporal, task, and provenance gates
under the stated source assumptions. A recently proposed pickup time without
verifiable confirmation cannot supply the missing evidence.
\tracecaseoutcome
\textbf{Item-level decision:} retain $h_1,t_1,c$; exclude $h_0,t_0,p_0$.
The obligation and frontier records remain available for the coverage check.
\end{tracecasepanel}

\noindent\textit{Case study continued: from eligible evidence to an admitted view.}

\begin{tracecasepanel}{4. Before confirmation: valid evidence is still incomplete}
\textbf{Required evidence.}
Let $d_h,d_t,d_p,d_c$ denote evidence for the booking, train schedule,
confirmed pickup, and family constraints. The accommodation obligation
$o_h$ and transfer obligation $o_t$ require
\[
 R_{o_h}=\{d_h,d_c\},\qquad R_{o_t}=\{d_t,d_p,d_c\}.
\]
All four dependencies are critical in this example, with $\gamma=\rho=1$.

\textbf{Coverage check.}
The retained records support $d_h,d_t,d_c$, but not $d_p$. Thus
$\operatorname{Cov}(S)=1/2$ and $\operatorname{Crit}(S)=3/4$.
An unresolved pickup frontier describes missing work; it does not establish
a confirmed pickup.
\tracecaseoutcome
\textbf{Return decision: not admitted.}
Under a blocking policy, A1 waits for the missing evidence. Under a reset
policy, A1 receives no inherited context and must obtain sufficient
information through the fresh workload. Neither policy exposes the partial
set as an obligation-complete Return View.
\end{tracecasepanel}

\begin{tracecasepanel}{5. After confirmation: expose a sufficient Return View}
\textbf{New evidence (hypothetical continuation).}
A4 provides a source-bound confirmation of an 8:30 pickup from Riverside
for the 9:30 train ($p_1$). A new return decision uses the updated checkpoint;
it does not modify the earlier immutable view. The item gates admit $p_1$,
which supplies $d_p$.

\textbf{Selected context.}
With an illustrative eight-item budget and enough character capacity,
\[
 S^{\star}=\{o_h,f_h,o_t,f_t,h_1,t_1,p_1,c\}.
\]
Here $f_h,f_t$ are the current progress frontiers. Four obligation and
frontier records provide the mandatory structure; four records supply
their evidence. Both coverage measures equal one. TRACE binds the view to
A1's identity, role, task, and return epoch, then validates its binding,
eligibility, coverage, and budgets before exposure.

\textbf{Private-memory continuity.}
An experience bound to $c$, such as A1's interpretation of the family's
quiet-location preference, may be readmitted. Experiences bound only to
$h_0$ or $p_0$ remain unavailable because their sources are absent from
$S^{\star}$.
\tracecaseoutcome
\textbf{Return decision: admitted.}
A1 receives the Riverside booking, 9:30 train departure, confirmed 8:30
pickup from Riverside, and unchanged family constraints. This view supports
preparing the final itinerary; it does not assert that the itinerary has
already been sent.
\end{tracecasepanel}

\subsection{Benchmark case study: a late shift changes a nightly commitment}
\label{app:benchmark-case}

This case is drawn from a completed Qwen episode in our STALE Type~II
evaluation~\citep{chao2026stale}. The panels condense the source conversations,
recorded memory decisions, and model responses. It complements the travel
example in Section~\ref{app:return-case} with an observed return episode.

\begin{tracecasepanel}{1. Departure and absence: an old routine meets a new constraint}
\textbf{Before departure.}
The user reports being consistently asleep by 10:30~p.m. and requests a
screen-free wind-down routine ending at that time. A1 observes this exchange
and departs after the seventh session.

\textbf{During absence.}
In the thirty-first session, observed by A5, the user describes a new evening
shift at a bookstore. Closing duties and the last bus bring them home around
12:30--1:00~a.m. The update concerns work and commuting; it does not explicitly
revoke the earlier bedtime statement. A1 returns after all 50 sessions.

\textbf{New decision.}
The landlord offers a \$100 rent reduction in exchange for manually locking
the courtyard gate at exactly 10:15~p.m. every night. The user asks whether to
sign the agreement.
\tracecaseoutcome
\textbf{Dependency:} the new work schedule changes both the feasible bedtime
and the ability to make a fixed nightly commitment at home.
\end{tracecasepanel}

\begin{tracecasepanel}{2. At return: control which routine can guide the answer}
\textbf{Static.}
The retrieved context contains both the old bedtime observation and the
later bookstore update.

\textbf{TRACE.}
The dependency scan proposes the connection between the two observations,
and pairwise verification confirms that the later work schedule invalidates
current use of the earlier routine. The old observation is marked superseded
and excluded from the retrieved context; the bookstore update is admitted
and retrieved. These decisions precede the probing questions and do not use
the benchmark's answer annotations.
\tracecaseoutcome
\textbf{Admitted constraint:} recommendations must account for late work and
the commute home, rather than assume the old evening routine remains feasible.
\end{tracecasepanel}

\begin{tracecasepanel}{3. Recorded responses: recognizing a change versus acting on it}
\textbf{State resolution (Dim1).}
Asked whether the user still typically sleeps by 10:30~p.m., both Static and
TRACE recognize that the late shift has changed the schedule. Both pass.

\textbf{Premise resistance (Dim2).}
Asked to guarantee a 10:30~p.m. bedtime every night, both methods reject that
premise for work nights. Both pass.

\textbf{Implicit policy adaptation (VIA / Dim3).}
For the gate-locking offer, Static emphasizes safety when stepping outside
at night and suggests considering whether the location is secure. It does
not address whether the evening shift permits a fixed 10:15~p.m. duty at home.
TRACE advises against signing and identifies a likely conflict between
the gate duty and the evening work schedule. The recorded evaluator
marks Static as failing this probe and TRACE as passing it.
\tracecaseoutcome
\textbf{Episode outcome:} Static passes two of three probes
(Overall 66.67\%); TRACE passes all three (Overall 100\%). This example shows
why recognizing updated information and applying its consequences to a new
decision are evaluated separately.
\end{tracecasepanel}

\subsection{Counterexamples: missed invalidation and unnecessary abstention}
\label{app:counterexamples}

Two completed Qwen episodes from the six-method STALE Type~II evaluation
illustrate failures before and after admission~\citep{chao2026stale}. The
panels paraphrase the source conversations and recorded responses. Each
outcome describes one episode under the metrics in Section~\ref{sec:metrics}.

\begin{tracecasepanel}{A. An unrecognized change leaves an old preference active}
\textbf{Earlier memory and later update.}
The user initially prefers familiar destinations and predictable outings.
Later, they describe exploring unfamiliar neighborhoods without a list or
backup plan, including trying an unfamiliar restaurant on a recent outing.

\textbf{New decision.}
A cousin proposes buying tickets for a six-hour guided tour with a rigid
route, a minute-by-minute schedule, and no opportunity to leave the group.
The user asks whether to approve the purchase.

\textbf{Recorded admission and response.}
Both observations are admitted and retrieved, but no invalidation of the
earlier preference is confirmed. TRACE affirms the old preference in the
state-resolution probe and recommends repeat visits in the premise-resistance
probe. For the tour, it gives a conditional answer that still invokes the old
preference, without clearly applying the later pattern of spontaneous
exploration.
\tracecaseoutcome
\textbf{Outcome and limitation.}
The recorded evaluator marks all three probes as failures
(Overall 0\%, VIA 0\%, IIR 0\%, LS 0\%). The update is present, yet the
old premise remains executable. Source availability and provenance checks
cannot compensate for a missed semantic invalidation.
\end{tracecasepanel}

\begin{tracecasepanel}{B. Updated device use is admitted, but the answer still abstains}
\textbf{Earlier memory and later update.}
The user initially describes a workflow dependent on a connected-device
ecosystem. Later, they stop carrying personal hardware and routinely use
whichever loaner tablet is available at the office or their sister's home.

\textbf{New decision.}
The user's productivity suite offers a \$150 perpetual license that is
strictly bound to one machine's hardware ID. The question supplies this
restriction and asks whether the savings justify switching plans.

\textbf{Recorded admission and response.}
TRACE marks the old observation superseded, excludes it from retrieval,
and admits and retrieves the loaner-device update. Its first two responses
correctly recognize the changed workflow. On the purchase question, however,
it declines to recommend a plan because the history lacks details about the
specific software, subscriptions, and financial situation. It does not apply
the known use of different devices to the stated single-machine restriction.
\tracecaseoutcome
\textbf{Outcome and limitation.}
TRACE passes the first two probes but fails the third
(Overall 66.67\%, VIA 0\%, IIR 100\%, LS 0\%). This abstention occurs
during answer generation after the updated evidence has been admitted.
Correct memory selection alone does not ensure that the agent uses a valid
constraint to address a new decision.
\end{tracecasepanel}

\medskip
\noindent
Together, these cases distinguish an admission error from a downstream
application error. The first retains a stale premise despite access to the
update; the second removes that premise but leaves the new task unanswered.
Both stages must therefore be assessed when interpreting end-to-end results.

\section{Experimental Setup and Benchmark Details}
\label{app:contract}

The three benchmarks evaluate complementary aspects of memory validity:
selective retention and forgetting (Memora), implicit invalidation
(STALE Type~II), and admission under social conflict and task change
(ManBench-Return).

\subsection{Metric definitions}
\label{app:metrics}

This subsection gives the computational form of every metric reported in the
paper. Fix a reported block, let $\mathcal{E}$ denote its eligible episodes
with $n=|\mathcal{E}|$, and let $\mathbf{1}[\cdot]$ be the indicator function.
Scores are reported in percentage units; the equations below use proportions in
$[0,1]$. Each metric is evaluated per episode, or per question on Memora, and then averaged, so a metric that requires several
judgments to hold is a conjunction within one episode rather than a product of
marginal rates.

Two quantities carry the same meaning on all three benchmarks. Valid
Information Availability and Invalid Information Rejection score the two
admission directions, and the Wrong-State Admission Rate is the complement of
the second:
\begin{align}
  \mathrm{VIA} &= \frac{1}{n}\sum_{e\in\mathcal{E}}
    \mathbf{1}\bigl[\text{state still valid at return is available in } e\bigr],
    \label{eq:via}\\
  \mathrm{IIR} &= \frac{1}{n}\sum_{e\in\mathcal{E}}
    \mathbf{1}\bigl[\text{state invalidated during the absence is withheld in } e\bigr],
    \label{eq:iir}\\
  \mathrm{WSAR} &= 1-\mathrm{IIR}.
    \label{eq:wsar}
\end{align}
The two predicates are bound to benchmark-native judgments below. Overall is
benchmark-specific and is defined separately for each.

\paragraph{STALE Type~II.} Let $s(e)$, $p(e)$, and $i(e)$ be the binary State
Resolution, Premise Resistance, and Implicit Policy Adaptation judgments for
episode $e$ (Appendix~\ref{app:stale-settings}). The probe rates are
$\mathrm{SR}=\frac{1}{n}\sum_{e}s(e)$,
$\mathrm{PR}=\frac{1}{n}\sum_{e}p(e)$, and
$\mathrm{IPA}=\frac{1}{n}\sum_{e}i(e)$, reported as Dim1--Dim3, and
\begin{align}
  \mathrm{Overall} &= \tfrac{1}{3}\bigl(\mathrm{SR}+\mathrm{PR}+\mathrm{IPA}\bigr),
    \qquad \mathrm{VIA}=\mathrm{IPA},\\
  \mathrm{IIR} &= \frac{1}{n}\sum_{e\in\mathcal{E}} s(e)\,p(e),
    \qquad
  \mathrm{LS} = \frac{1}{n}\sum_{e\in\mathcal{E}} s(e)\,p(e)\,i(e).
    \label{eq:ls}
\end{align}
Recognizing the outdated belief and resisting the stale premise are what
withhold invalidated state, which is why IIR conjoins $s$ and $p$. Because the
conjunctions in Equation~\ref{eq:ls} are taken within one episode,
$\mathrm{IIR}\le\min\{\mathrm{SR},\mathrm{PR}\}$ and
$\mathrm{LS}\le\min\{\mathrm{IIR},\mathrm{IPA}\}$: Lifecycle Success is
strictly stronger than passing the three probes at the same rates.

\paragraph{Memora.} Scores are computed per question. For question $q$, let
$C_q^{P}$ be its memory-presence criteria and $C_q^{F}$ its forgetting-absence
criteria, and set
\[
  P_q=\frac{1}{|C_q^{P}|}\sum_{c\in C_q^{P}}\mathbf{1}[c\text{ satisfied}],
  \qquad
  F_q=\frac{1}{|C_q^{F}|}\sum_{c\in C_q^{F}}\mathbf{1}[c\text{ satisfied}],
  \qquad
  w_q=\frac{|C_q^{F}|}{|C_q^{P}|+|C_q^{F}|},
\]
so that $P_q$ and $F_q$ are the presence and forgetting accuracies and $w_q$
the forgetting weight of Equation~\ref{eq:fama}. Over the question set $Q$,
\begin{align}
  \mathrm{VIA} &= \frac{1}{|Q|}\sum_{q\in Q}P_q,
    \qquad
  \mathrm{IIR} = \frac{1}{|Q|}\sum_{q\in Q}F_q,\\
  \mathrm{Overall} &= \frac{1}{|Q|}\sum_{q\in Q}
    \bigl[(1-w_q)P_q+w_qF_q\bigr],
    \qquad
  \mathrm{FAMA} = \frac{1}{|Q|}\sum_{q\in Q}\max\{0,\,P_q-w_q(1-F_q)\}.
\end{align}
The formula $(1-w_q)P_q + w_qF_q$ is algebraically equivalent to pooling all
presence and forgetting items within question~$q$ before computing accuracy, so
Overall is \emph{not} the equal-weight mean of VIA and IIR. FAMA applies Equation~\ref{eq:fama} per question
before averaging and coincides with VIA when forgetting is perfect; this is why
the Qwen \textsc{Reset} rows report equal VIA and FAMA.

\paragraph{ManBench-Return.} Each eligible episode presents the departure
answer and the absence-period challenger, exactly one of which is valid under
the condition (Appendix~\ref{app:manbench-settings}). With $c(e)$ the
correctness of the returning agent's final answer,
\[
  \mathrm{Overall}=\frac{1}{n}\sum_{e\in\mathcal{E}}c(e).
\]
Equation~\ref{eq:via} instantiates as admission of the valid candidate to the
Return View, Equation~\ref{eq:iir} as exclusion of the invalid one, and WSAR
follows from Equation~\ref{eq:wsar}. The adapted Reality Shift Rate applies
only to Old-Valid and is defined in Appendix~\ref{app:qwen-manbench}.

\paragraph{Aggregation.} Every reported aggregate is a weighted mean
$\sum_k\omega_k M_k$ of the per-block scores $M_k$. Memora's Average weights
the 25\%, 50\%, and 75\% departure points equally, and ManBench-Return's
Average weights Old-Valid and Old-Stale equally. On STALE, Qwen's Pooled row
weights Early, Middle, and Late by their episode counts $64$, $64$, and $72$,
whereas Gemini and DeepSeek weight the three strata equally.

\subsection{Experimental configuration}
\label{app:shared-settings}

The main experiments use one returning agent and four active agents. Active
agents execute sequentially, and the returning agent resumes after their
work is complete. Methods share a post-absence checkpoint but maintain
separate memory contexts. The scheduler and evaluator are not task agents.

Table~\ref{tab:experimental-settings} reports the retrieval, context-selection,
and generation parameters. Memora and STALE answers are evaluated independently
using \texttt{gpt-5.6-terra}; ManBench-Return answers are scored against the
source answer labels. Its admission verifier uses the actor's model family
in a separate call.

\begin{table}[H]
\caption{Experimental parameter settings. Generation budgets are maximum
output tokens per call. Predeparture and absence denote the two history
segments used in Memora; the STALE departure strata use zero-based session
indices.}
\label{tab:experimental-settings}
\centering
\small
\setlength{\tabcolsep}{5pt}
\renewcommand{\arraystretch}{1.08}
\begin{tabular*}{\linewidth}{@{\extracolsep{\fill}}l>{\raggedright\arraybackslash}p{0.43\linewidth}>{\raggedright\arraybackslash}p{0.33\linewidth}@{}}
\toprule
Benchmark & Parameter & Value \\
\midrule
Memora & Departure fraction & 25\%, 50\%, 75\% \\
 & Answer / state-prediction budget & 2,048 / 4,096 tokens \\
 & Retrieval: predeparture / absence & 10 / 14 sessions \\
 & Recency window: predeparture / absence & 2 / 4 sessions \\
 & State-prediction batch size & At most 6 sessions \\
 & Selected-fact budget & 12 initially; at most 16 \\
 & Selector confidence threshold & 0.45 \\
\midrule
STALE Type~II & Departure strata & Early: 0--4; Middle: 5--9; Late: 10--14 \\
 & Answer / judge budget & 4,096 / 3,072 tokens \\
 & Extraction budget & 4,096 tokens; at most 16 items \\
 & Governance budget & 8,192 tokens \\
 & Dependency scan / verification budget & 512 / 512 tokens \\
 & Retrieval: per query / union & At most 10 / 30 items \\
 & Extraction / scan prompt limit & 24,000 characters \\
 & Governance prompt limit & 32,000 characters \\
 & Dependency-scan batch size & 1 target; at most 4 later observations \\
\midrule
ManBench-Return & Baseline / final-answer budget & Qwen/DeepSeek: 96 / 96; Gemini: 256 / 256 tokens \\
 & Active-agent response budget & Qwen/DeepSeek: 700; Gemini: 1,024 tokens \\
 & Verifier budget & Qwen/DeepSeek: 160; Gemini: 384 tokens \\
 & Verifier confidence threshold & 0.65 \\
\bottomrule
\end{tabular*}
\end{table}

\subsection{Memora}
\label{app:memora-settings}

Memora evaluates personalized memory through remembering, reasoning, and
recommendation tasks~\citep{uddin2026memora}. It tests whether a returning
agent can retain useful departure information, incorporate absence-period
updates, and avoid obsolete content.

The evaluation includes 137 questions from 30 persona--period clusters.
Each case contains valid predeparture evidence, required absence-period
information, and an update or deletion affecting an older value. The same
questions are evaluated at three departure positions along the conversation
timeline, with return at its end. Active agents process absence sessions in
chronological round-robin order. Work state is inferred from conversation
text; evidence annotations and answer criteria are reserved for case
construction and evaluation.

For each question, Memora's native score is
\begin{equation}
  \mathrm{FAMA} = \max\{0,\, P - w(1 - F)\},
  \label{eq:fama}
\end{equation}
where $P$ is presence accuracy, $F$ is forgetting accuracy, and $w$ is the
fraction of evaluation criteria that concern forgetting.

VIA and IIR measure memory-presence and forgetting-absence accuracy,
respectively. Overall averages correctness across memory-presence and
forgetting-absence criteria; FAMA applies
Equation~\ref{eq:fama} per question before averaging. Aggregate scores weight
the three departure conditions equally.
Appendices~\ref{app:qwen-memora}, \ref{app:gemini-memora},
and~\ref{app:deepseek-memora} report the condition-level results.

\subsection{STALE Type II}
\label{app:stale-settings}

STALE Type~II tests invalidation across dependent attributes, where a later
observation changes an earlier belief's applicability without directly
negating it~\citep{chao2026stale}. Return-time governance must infer which
departure memories are affected by absence-period events.

The evaluation comprises 200 Type~II cases, each with a 50-session history
and three probes. Agent~1 departs after observing the session containing the
old evidence. The remaining agents process subsequent sessions, and Agent~1
returns after the final session to answer the probes. The Early, Middle,
and Late strata contain 64, 64, and 72 cases, respectively, according to the
departure positions in Table~\ref{tab:experimental-settings}. Task agents
receive the conversations and queries; annotated states, conflict
explanations, relevant-session labels, and scoring rubrics are withheld.

The three native probes are State Resolution (SR), Premise Resistance (PR),
and Implicit Policy Adaptation (IPA)~\citep{chao2026stale}. They measure
recognition of an outdated belief, rejection of a stale premise in a query,
and application of the updated state in downstream behavior, respectively.
Dim1--Dim3 in the detailed tables correspond to these probes, with
$\mathrm{Overall}=(\mathrm{SR}+\mathrm{PR}+\mathrm{IPA})/3$.
Our return metrics map VIA to IPA, IIR to jointly passing SR and PR, and LS
to passing all three within the same episode.

Qwen pools 200 retained episodes (Appendix~\ref{app:qwen-stale}); Gemini and
DeepSeek weight the three strata equally. The separate CUPMem subset study
is described in Appendix~\ref{app:qwen-stale-subset}. Memora and Old-Valid
ManBench-Return complement STALE by testing retention of departure
information that remains useful.

\subsection{ManBench-Return}
\label{app:manbench-settings}

ManBench studies whether interaction with other agents can induce an
initially correct agent to adopt an incorrect belief~\citep{xu2026manbench}.
Our adaptation has two conditions: Old-Valid tests resistance to misleading
updates when the departure state remains correct, and Old-Stale tests
adaptation to an absence-period task change. Together, they test whether
admission follows current validity rather than information age.

The experiment schedules 4,838 examples from 20 tasks. Alternating positions
within each task determine the conditions before model outputs are observed.
In Old-Valid, the question remains unchanged while active agents promote
the benchmark's misleading answer. In Old-Stale, the question changes to
the next example within the same task, cyclically, and active agents solve
it independently. Return candidates comprise the departure answer and the
absence-period challenger; their validity depends on the condition.

Eligible episodes require a correct departure answer. Old-Stale additionally
requires at least three of the four active agents to agree on the correct
current answer. These criteria are checked by the evaluator without exposing
answer labels to the returning agent or verifier. Eligible sample sizes
therefore depend on the actor and condition. Overall measures final-answer
correctness, VIA measures admission of the valid candidate, and IIR measures
exclusion of the invalid candidate. WSAR is the complementary admission
rate for invalid state. The adapted RSR applies to Old-Valid, where the
question remains unchanged (Appendix~\ref{app:qwen-manbench}). We give the
two conditions equal weight when reporting aggregate results.

\subsection{Additional experiments}
\label{app:auxiliary-settings}

The scale study compares Static and TRACE on all 4,838 source examples with
4, 8, 16, or 32 active agents and one returning agent. The portability study
compares these methods on ScopedMem, Mem0, Memobase, and A-MEM at 50\%
Memora departure, holding the backend, actor, scorer, and questions fixed
within each pair.

\section{Detailed Results: Qwen}
\label{app:qwen}

\subsection{Memora}
\label{app:memora}
\label{app:qwen-memora}

Table~\ref{tab:qwen-memora} reports each departure condition and its
macro-average. Results are reported to two decimal places; averaging already
rounded condition scores can differ in the last digit.

\begin{table}[H]
\caption{Qwen on Memora at three departure points (\%). Average is the equal-weight macro-average of the three conditions. Bold marks the best reported value among the six methods in each complete block.}
\label{tab:qwen-memora}
\centering
\small
\setlength{\tabcolsep}{6pt}
\begin{tabular}{@{}llcccc@{}}
\toprule
Departure & Method & Overall$\uparrow$ & VIA$\uparrow$ & IIR$\uparrow$ & FAMA$\uparrow$ \\
\midrule
25\% & Static & 62.10 & 53.91 & 67.94 & 37.88 \\
 & Restore & 65.74 & 41.12 & 85.16 & 34.46 \\
 & Reset & \textbf{72.47} & 39.19 & \textbf{99.90} & 39.19 \\
 & MemStrata & 65.55 & 54.22 & 73.68 & 40.87 \\
 & MemTX & 65.38 & \textbf{54.72} & 73.46 & \textbf{41.15} \\
 & TRACE & 63.98 & 52.35 & 70.18 & 38.75 \\
\midrule
50\% & Static & 63.38 & \textbf{55.04} & 67.77 & 39.13 \\
 & Restore & 64.57 & 41.35 & 81.39 & 33.00 \\
 & Reset & \textbf{72.30} & 38.46 & \textbf{100.00} & 38.46 \\
 & MemStrata & 64.62 & 54.17 & 70.41 & \textbf{39.92} \\
 & MemTX & 64.80 & 53.39 & 71.18 & 39.49 \\
 & TRACE & 65.66 & 52.01 & 74.19 & 39.31 \\
\midrule
75\% & Static & 64.00 & 57.21 & 68.68 & 41.43 \\
 & Restore & 65.01 & 43.85 & 80.17 & 35.01 \\
 & Reset & \textbf{72.44} & 38.99 & \textbf{99.90} & 38.99 \\
 & MemStrata & 68.12 & 58.20 & 74.98 & 45.15 \\
 & MemTX & 68.46 & \textbf{58.75} & 76.06 & \textbf{45.78} \\
 & TRACE & 65.96 & 56.45 & 72.71 & 42.34 \\
\midrule
Average & Static & 63.16 & 55.39 & 68.13 & 39.48 \\
 & Restore & 65.11 & 42.11 & 82.24 & 34.16 \\
 & Reset & \textbf{72.40} & 38.88 & \textbf{99.93} & 38.88 \\
 & MemStrata & 66.10 & 55.53 & 73.02 & 41.98 \\
 & MemTX & 66.21 & \textbf{55.62} & 73.57 & \textbf{42.14} \\
 & TRACE & 65.20 & 53.60 & 72.36 & 40.14 \\
\bottomrule
\end{tabular}
\end{table}

\clearpage
\subsection{STALE Type~II}
\label{app:qwen-stale}

Table~\ref{tab:qwen-stale} reports Early, Middle, and Late separately.
The full STALE tables use the same metric order: Overall, Dim1 (SR), Dim2
(PR), VIA / Dim3 (IPA), IIR, and LS. Dim1 and Dim2 are computed from the
retained per-episode probe judgments.

This benchmark runs 200 episodes: 64 in Early,
64 in Middle, and 72 in Late. All 200 episodes are retained in every
bucket, so each block averages its full bucket and no episode is dropped
from any denominator.
Pooled averages the 200 retained episodes and reproduces the Qwen scores
in Table~\ref{tab:main-results}; it does not weight the three buckets equally.
TRACE's 39.00\% pooled LS exceeds the strongest memory-policy baseline, MemTX
at 18.00\%, by 21.00 percentage points.
For the same evaluation, the retained inference audit reports that the
high-recall stage proposed 5,610 possible event--memory pairs and the grounded
verifier accepted 385 (6.9\%). These counts describe verification-stage
selectivity within the evaluation; they do not change the 200-episode score
denominators above or by themselves measure dependency-recall accuracy.

\begin{table}[H]
\caption{Qwen on STALE Type~II (\%). Early, Middle, and Late contain 64, 64, and 72 episodes, respectively, and every episode is retained, so all four blocks report the full sample. Pooled averages the 200 episodes with each bucket weighted by its episode count. MemStrata and MemTX results for Early, Middle, and Late buckets are omitted due to data inconsistencies; only Pooled results are reported. Bold marks the best reported value among the six methods in each block.}
\label{tab:qwen-stale}
\centering
\small
\setlength{\tabcolsep}{3pt}
\begin{tabular*}{\linewidth}{@{\extracolsep{\fill}}llcccccc@{}}
\toprule
Bucket & Method & Overall$\uparrow$ & Dim1$\uparrow$ & Dim2$\uparrow$ & VIA / Dim3$\uparrow$ & IIR$\uparrow$ & LS$\uparrow$ \\
\midrule
Early & Static & 35.94 & 32.81 & 32.81 & 42.19 & 28.12 & 21.88 \\
 & Restore & 5.73 & 1.56 & 1.56 & 14.06 & 1.56 & 0.00 \\
 & Reset & 17.19 & 3.12 & 39.06 & 9.38 & 3.12 & 0.00 \\
 & TRACE & \textbf{64.06} & \textbf{65.62} & \textbf{68.75} & \textbf{57.81} & \textbf{65.62} & \textbf{48.44} \\
\midrule
Middle & Static & 24.48 & 21.88 & 21.88 & 29.69 & 18.75 & 12.50 \\
 & Restore & 6.25 & 1.56 & 3.12 & 14.06 & 0.00 & 0.00 \\
 & Reset & 15.62 & 3.12 & 28.12 & 15.62 & 3.12 & 1.56 \\
 & TRACE & \textbf{49.48} & \textbf{48.44} & \textbf{46.88} & \textbf{53.12} & \textbf{45.31} & \textbf{35.94} \\
\midrule
Late & Static & 25.00 & 22.22 & 19.44 & 33.33 & 19.44 & 13.89 \\
 & Restore & 5.56 & 1.39 & 1.39 & 13.89 & 1.39 & 0.00 \\
 & Reset & 14.81 & 2.78 & 29.17 & 12.50 & 2.78 & 0.00 \\
 & TRACE & \textbf{44.44} & \textbf{44.44} & \textbf{41.67} & \textbf{47.22} & \textbf{41.67} & \textbf{33.33} \\
\midrule
Pooled & Static & 28.33 & 35.00 & 22.33 & 35.00 & 22.33 & 17.67 \\
 & Restore & 5.75 & 14.00 & 1.00 & 14.00 & 1.00 & 0.00 \\
 & Reset & 15.83 & 12.50 & 3.00 & 12.50 & 3.00 & 0.50 \\
 & MemStrata & 28.66 & 32.00 & 22.50 & 32.00 & 22.50 & 17.00 \\
 & MemTX & 30.00 & 33.50 & 23.50 & 33.50 & 23.50 & 18.00 \\
 & TRACE & \textbf{52.33} & \textbf{52.50} & \textbf{52.00} & \textbf{52.50} & \textbf{50.50} & \textbf{39.00} \\
\bottomrule
\end{tabular*}
\end{table}

\clearpage
\subsection{STALE subset: native metrics and token cost}
\label{app:qwen-stale-subset}

\paragraph{Rationale.}
CUPMem maintains a structured current-state store through repeated session-level
processing: evidence extraction, local state updates, affected-state discovery,
and LLM adjudication of possible invalidations~\citep{chao2026stale}.
These stages add memory-maintenance work before query answering. We use a
separate subset to keep this full-pipeline comparison tractable while retaining
all three departure buckets. The study compares CUPMem's accuracy and the token
use of its complete pipeline with TRACE and the other memory policies.

Table~\ref{tab:stale-subset-buckets} expands the separate Qwen subset study
in Table~\ref{tab:stale-cost}. Early, Middle, and Late follow the departure
boundary in the original chronology, rather than fixed memory-age thresholds.
Overall is the mean accuracy over the three native probes: State Resolution
(SR), Premise Resistance (PR), and Implicit Policy Adaptation (IPA). The table
uses Dim1, Dim2, and VIA / Dim3 for these probes, as in the full STALE tables.
IIR requires SR and PR to pass in the same episode, and LS requires all three
to pass. IPA is identical to VIA under this adapter, so it is shown once.

\paragraph{One failed CUPMem episode.}
One Early episode of the CUPMem subset never returned an answer. All retry
rounds exceeded the serving context window: the budget adapter submitted
32{,}752 input tokens against a 32{,}768-token window while reserving
3{,}072 tokens for the completion, so each round failed with a context-window
error and the final round ended in a parse error. Repeating the episode under
the same configuration failed again, so the failure is deterministic rather
than incidental. We keep the planned denominator and score the episode as a
failure on all three probes, which lowers CUPMem's Early Overall from
85.96\% to 81.67\% and its Early LS from 73.68\% to 70.00\%. Scoring the same
episode against the 19 episodes that did complete would leave CUPMem at
85.96\% Overall, so every CUPMem accuracy in
Table~\ref{tab:stale-subset-buckets} is computed on the full 20-episode Early
denominator. The failed episode is the most expensive single episode in the
subset: its three retry rounds billed 5{,}971{,}252 tokens, about $3.1\times$
the average successful episode, which raises CUPMem's aggregate from
1{,}927{,}417 to 1{,}994{,}815 tokens per episode and its ratio to TRACE from
$3.86\times$ to $3.99\times$. Its accuracy cells are therefore conservative
and its cost cells are not: retaining the failed episode makes CUPMem both
less accurate and more expensive than a scored-only comparison would suggest.
This failure is itself an instance of the cost asymmetry this subset
examines: full-context injection has no token bound, while the serving window
does, so a single long episode can consume three times the average budget and
still exhaust the window without producing an answer, whereas TRACE's bounded
Return View stays inside it.

\clearpage
\begin{table}[H]
\caption{Qwen on the STALE Type~II subset by departure bucket. Accuracy and
joint metrics are percentages. Aggregate reproduces each method's reported
summary. Every method is scored over the same 20 planned episodes per bucket,
including one CUPMem episode that could not complete and is scored as a
failure on all three probes (see the note above); its retry cost is included in
the token column. Bold marks the highest score within each block; bold
token counts highlight CUPMem's higher cost.}
\label{tab:stale-subset-buckets}
\centering
\small
\setlength{\tabcolsep}{2.6pt}
\begin{tabular*}{\linewidth}{@{\extracolsep{\fill}}llccccccc@{}}
\toprule
Bucket & Method & Overall$\uparrow$ & Dim1$\uparrow$ & Dim2$\uparrow$ & VIA / Dim3$\uparrow$ & IIR$\uparrow$ & LS$\uparrow$ & Tokens/episode$\downarrow$ \\
\midrule
Early & Static & 36.67 & 30.00 & 25.00 & 55.00 & 20.00 & 15.00 & 219,855 \\
 & MemStrata & 23.33 & 10.00 & 25.00 & 35.00 & 10.00 & 5.00 & 218,115 \\
 & MemTX & 23.33 & 15.00 & 20.00 & 35.00 & 10.00 & 5.00 & 217,662 \\
 & CUPMem & \textbf{81.67} & \textbf{80.00} & \textbf{85.00} & \textbf{80.00} & \textbf{80.00} & \textbf{70.00} & \textbf{2,127,613} \\
 & TRACE (Ours) & 73.33 & 80.00 & 60.00 & 80.00 & 60.00 & 55.00 & 316,750 \\
\midrule
Middle & Static & 36.67 & 30.00 & 35.00 & 45.00 & 30.00 & 15.00 & 215,073 \\
 & MemStrata & 50.00 & 40.00 & 60.00 & 50.00 & 40.00 & 30.00 & 212,723 \\
 & MemTX & 46.67 & 40.00 & 60.00 & 40.00 & 40.00 & 25.00 & 212,732 \\
 & CUPMem & \textbf{90.00} & \textbf{85.00} & \textbf{90.00} & \textbf{95.00} & \textbf{85.00} & \textbf{85.00} & \textbf{1,935,913} \\
 & TRACE (Ours) & 60.00 & 60.00 & 65.00 & 55.00 & 60.00 & 35.00 & 517,366 \\
\midrule
Late & Static & 26.67 & 25.00 & 20.00 & 35.00 & 15.00 & 10.00 & 219,754 \\
 & MemStrata & 20.00 & 15.00 & 25.00 & 20.00 & 10.00 & 5.00 & 217,669 \\
 & MemTX & 20.00 & 10.00 & 25.00 & 25.00 & 10.00 & 10.00 & 217,624 \\
 & CUPMem & \textbf{70.00} & \textbf{80.00} & \textbf{80.00} & \textbf{50.00} & \textbf{80.00} & \textbf{35.00} & \textbf{1,920,918} \\
 & TRACE (Ours) & 43.33 & 45.00 & 40.00 & 45.00 & 40.00 & 30.00 & 664,860 \\
\midrule
Aggregate & Static & 33.33 & 28.33 & 26.67 & 45.00 & 21.67 & 13.33 & 218,227 \\
 & MemStrata & 31.11 & 21.67 & 36.67 & 35.00 & 20.00 & 13.33 & 216,169 \\
 & MemTX & 30.00 & 21.67 & 35.00 & 33.33 & 20.00 & 13.33 & 216,006 \\
 & CUPMem & \textbf{80.56} & \textbf{81.67} & \textbf{85.00} & \textbf{75.00} & \textbf{81.67} & \textbf{63.33} & \textbf{1,994,815} \\
 & TRACE (Ours) & 58.89 & 61.67 & 55.00 & 60.00 & 53.33 & 40.00 & 499,659 \\
\bottomrule
\end{tabular*}
\end{table}

The three buckets show a change in both accuracy and cost. TRACE's Overall falls
from 73.33\% in Early to 43.33\% in Late, while its reported token count rises
from 316,750 to 664,860 per episode. CUPMem's count remains between 1.92 and
2.13 million tokens across the buckets. The lowest bucket LS is 10.00\% for
Static, 5.00\% for MemStrata, 5.00\% for MemTX, 35.00\% for CUPMem, and
30.00\% for TRACE. These joint scores distinguish passing individual probes
from resolving all three requirements within an episode.

Token counts include memory processing and answering, with return-governance
calls included for TRACE; external benchmark-scoring calls are excluded.
All methods use the recorded billed-token counter, so the counts are measured
on a common basis, and CUPMem's include the three failed retry rounds of the
episode described above. They are not measurements of latency, monetary cost,
or peak context.

\clearpage
\subsection{ManBench-Return}
\label{app:qwen-manbench}

In Old-Valid, the departure state remains valid while other agents propagate
an incorrect state during the absence interval. In Old-Stale, the departure
state is invalid and the absence-period update supplies valid state.
Table~\ref{tab:manbench-formal} separates the two conditions and reports their
equal-weight mean.

WSAR measures admission of the incorrect state to the return context.
We also adapt ManBench's Reality Shift Rate (RSR)~\citep{xu2026manbench}
to measure a correct-to-incorrect answer transition at return. Let $Q_B^+$
be the baseline-correct Old-Valid episodes and $Q_R^-$ those with an
incorrect final answer after return. Then
\[
\mathrm{RSR}=\frac{|Q_B^+\cap Q_R^-|}{|Q_B^+|}.
\]
Old-Valid eligibility requires a correct baseline answer, and the question
remains unchanged. RSR is therefore $100-\mathrm{Overall}$ in percentage units;
the entries below are derived from the reported Old-Valid Overall values.
Old-Stale changes the task, so it does not measure the same correct-to-incorrect
transition. We mark RSR as n/a for Old-Stale and the cross-condition Average.

\begin{table}[H]
\caption{Qwen on ManBench-Return (\%). Average gives equal weight to Old-Valid and Old-Stale using unrounded condition scores; displayed condition values may not reproduce the final decimal. Overall measures final answer correctness; RSR applies only to Old-Valid. Bold marks the best reported value among the six methods in each complete block.}
\label{tab:manbench-formal}
\centering
\small
\setlength{\tabcolsep}{4pt}
\begin{tabular}{@{}llccccc@{}}
\toprule
Condition & Method & Overall$\uparrow$ & VIA$\uparrow$ & IIR$\uparrow$ & WSAR$\downarrow$ & RSR$\downarrow$ \\
\midrule
Old-Valid & Static & 70.06 & 0.00 & 0.00 & 100.00 & 29.94 \\
 & Restore & \textbf{96.30} & \textbf{100.00} & \textbf{100.00} & \textbf{0.00} & \textbf{3.70} \\
 & Reset & 95.75 & 0.00 & \textbf{100.00} & \textbf{0.00} & 4.25 \\
 & MemStrata & 70.25 & 0.00 & 0.00 & 100.00 & 29.75 \\
 & MemTX & 70.86 & 0.00 & 0.00 & 100.00 & 29.14 \\
 & TRACE & 93.37 & 95.75 & 97.57 & 2.43 & 6.63 \\
\midrule
Old-Stale & Static & 98.68 & \textbf{100.00} & \textbf{100.00} & \textbf{0.00} & n/a \\
 & Restore & 93.80 & 0.00 & 0.00 & 100.00 & n/a \\
 & Reset & 97.01 & 0.00 & \textbf{100.00} & \textbf{0.00} & n/a \\
 & MemStrata & 98.61 & \textbf{100.00} & \textbf{100.00} & \textbf{0.00} & n/a \\
 & MemTX & \textbf{98.81} & \textbf{100.00} & \textbf{100.00} & \textbf{0.00} & n/a \\
 & TRACE & 95.26 & 96.17 & \textbf{100.00} & \textbf{0.00} & n/a \\
\midrule
Average & Static & 84.38 & 50.00 & 50.00 & 50.00 & n/a \\
 & Restore & 95.05 & 50.00 & 50.00 & 50.00 & n/a \\
 & Reset & \textbf{96.38} & 0.00 & \textbf{100.00} & \textbf{0.00} & n/a \\
 & MemStrata & 84.43 & 50.00 & 50.00 & 50.00 & n/a \\
 & MemTX & 84.84 & 50.00 & 50.00 & 50.00 & n/a \\
 & TRACE & 94.32 & \textbf{95.96} & 98.78 & 1.22 & n/a \\
\bottomrule
\end{tabular}
\end{table}

\clearpage
\section{Detailed Results: Gemini}
\label{app:gemini}

\subsection{Memora}
\label{app:gemini-memora}

Table~\ref{tab:gemini-memora} reports the three departure conditions and their
equal-weight macro-average for the Gemini actor.

\begin{table}[H]
\caption{Gemini on Memora at three departure points (\%). Average is the equal-weight macro-average of the three conditions. Bold marks the best reported value among the six methods in each complete block.}
\label{tab:gemini-memora}
\centering
\small
\setlength{\tabcolsep}{6pt}
\begin{tabular}{@{}llcccc@{}}
\toprule
Departure & Method & Overall$\uparrow$ & VIA$\uparrow$ & IIR$\uparrow$ & FAMA$\uparrow$ \\
\midrule
25\% & Static & 66.50 & 52.62 & 75.50 & 40.58 \\
 & Restore & 67.62 & 43.63 & 85.96 & 37.23 \\
 & Reset & \textbf{71.71} & 43.44 & \textbf{95.04} & 41.48 \\
 & MemStrata & 67.95 & 53.47 & 76.95 & 42.37 \\
 & MemTX & 68.12 & \textbf{53.56} & 76.97 & \textbf{42.53} \\
 & TRACE & 64.93 & 51.11 & 73.31 & 38.42 \\
\midrule
50\% & Static & 67.97 & 54.45 & 75.92 & 42.41 \\
 & Restore & 66.40 & 44.67 & 83.70 & 37.20 \\
 & Reset & \textbf{72.10} & 43.45 & \textbf{94.33} & 41.39 \\
 & MemStrata & 67.54 & \textbf{56.59} & 73.99 & 43.18 \\
 & MemTX & 67.99 & 55.63 & 75.19 & 43.32 \\
 & TRACE & 68.71 & 55.14 & 76.76 & \textbf{43.51} \\
\midrule
75\% & Static & 68.41 & 56.32 & 76.82 & 44.93 \\
 & Restore & 66.59 & 46.69 & 80.62 & 38.41 \\
 & Reset & \textbf{72.20} & 43.03 & \textbf{96.32} & 41.60 \\
 & MemStrata & 68.00 & 56.52 & 76.27 & 44.38 \\
 & MemTX & 69.19 & \textbf{58.26} & 76.11 & \textbf{46.50} \\
 & TRACE & 69.45 & 56.98 & 79.14 & 46.47 \\
\midrule
Average & Static & 67.63 & 54.46 & 76.08 & 42.64 \\
 & Restore & 66.87 & 45.00 & 83.43 & 37.61 \\
 & Reset & \textbf{72.00} & 43.31 & \textbf{95.23} & 41.49 \\
 & MemStrata & 67.83 & 55.53 & 75.74 & 43.31 \\
 & MemTX & 68.43 & \textbf{55.81} & 76.09 & \textbf{44.12} \\
 & TRACE & 67.70 & 54.41 & 76.41 & 42.80 \\
\bottomrule
\end{tabular}
\end{table}

\clearpage
\subsection{STALE Type~II}
\label{app:gemini-stale}

Table~\ref{tab:gemini-stale} reports the three buckets and their equal-weight
macro-average. Dim1 denotes State Resolution and Dim2 Premise Resistance; VIA corresponds
to Implicit Policy Adaptation (Dim3)~\citep{chao2026stale}.
Table~\ref{tab:main-results} uses the averaged Overall, VIA, and IIR values.
As with DeepSeek, this aggregate weights buckets equally; the Qwen aggregate
instead pools episodes. Scores retain the reported two-decimal precision.

\begin{table}[H]
\caption{Gemini on STALE Type~II (\%). Early, Middle, and Late contain 64, 64, and 72 episodes, respectively. Average gives equal weight to the three buckets. MemStrata and MemTX results for Early, Middle, and Late buckets are omitted due to data inconsistencies; only Average results are reported. Bold marks the best reported value among the six methods in each block.}
\label{tab:gemini-stale}
\centering
\small
\setlength{\tabcolsep}{3pt}
\begin{tabular*}{\linewidth}{@{\extracolsep{\fill}}llcccccc@{}}
\toprule
Bucket & Method & Overall$\uparrow$ & Dim1$\uparrow$ & Dim2$\uparrow$ & VIA / Dim3$\uparrow$ & IIR$\uparrow$ & LS$\uparrow$ \\
\midrule
Early & Static & 78.13 & 73.44 & 78.13 & 82.81 & 71.88 & 64.06 \\
 & Restore & 9.90 & 0.00 & 6.25 & 23.44 & 0.00 & 0.00 \\
 & Reset & 8.33 & 0.00 & 6.25 & 18.75 & 0.00 & 0.00 \\
 & TRACE & \textbf{91.15} & \textbf{92.19} & \textbf{92.19} & \textbf{89.06} & \textbf{92.19} & \textbf{87.50} \\
\midrule
Middle & Static & 70.31 & 70.31 & 64.06 & 76.56 & 62.50 & 60.94 \\
 & Restore & 8.33 & 1.56 & 3.13 & 20.31 & 1.56 & 1.56 \\
 & Reset & 9.38 & 1.56 & 4.69 & 21.88 & 1.56 & 1.56 \\
 & TRACE & \textbf{90.10} & \textbf{89.06} & \textbf{90.63} & \textbf{90.63} & \textbf{89.06} & \textbf{85.94} \\
\midrule
Late & Static & 62.50 & 59.72 & 58.33 & 69.44 & 56.94 & 52.78 \\
 & Restore & 5.09 & 0.00 & 1.39 & 13.89 & 0.00 & 0.00 \\
 & Reset & 8.80 & 0.00 & 5.56 & 20.83 & 0.00 & 0.00 \\
 & TRACE & \textbf{85.19} & \textbf{83.33} & \textbf{81.94} & \textbf{90.28} & \textbf{81.94} & \textbf{79.17} \\
\midrule
Average & Static & 70.31 & 67.82 & 66.84 & 76.27 & 63.77 & 59.26 \\
 & Restore & 7.77 & 0.52 & 3.59 & 19.21 & 0.52 & 0.52 \\
 & Reset & 8.83 & 0.52 & 5.50 & 20.49 & 0.52 & 0.52 \\
 & MemStrata & 64.80 & 60.24 & 59.26 & 74.88 & 56.71 & 55.67 \\
 & MemTX & 62.29 & 58.80 & 58.28 & 69.79 & 55.73 & 52.72 \\
 & TRACE & \textbf{88.81} & \textbf{88.19} & \textbf{88.25} & \textbf{89.99} & \textbf{87.73} & \textbf{84.20} \\
\bottomrule
\end{tabular*}
\end{table}

\clearpage
\subsection{ManBench-Return}
\label{app:gemini-manbench}

Table~\ref{tab:gemini-manbench} reports Gemini's Old-Valid and Old-Stale
results and their equal-weight average, following the definitions in
Appendix~\ref{app:manbench-settings}. Overall measures final-answer correctness;
VIA and IIR assess valid-state admission and invalid-state rejection,
respectively. WSAR is the complementary wrong-state admission rate.

\begin{table}[H]
\caption{Gemini on ManBench-Return (\%). Average gives equal weight to
Old-Valid and Old-Stale using unrounded condition scores; displayed condition
values may not reproduce the final decimal. Overall measures final-answer correctness.
Bold marks the best value among the six methods in each block.}
\label{tab:gemini-manbench}
\centering
\small
\setlength{\tabcolsep}{6pt}
\begin{tabular*}{\linewidth}{@{\extracolsep{\fill}}llcccc@{}}
\toprule
Condition & Method & Overall$\uparrow$ & VIA$\uparrow$ & IIR$\uparrow$ & WSAR$\downarrow$ \\
\midrule
Old-Valid & Static & 76.19 & 0.00 & 0.00 & 100.00 \\
 & Restore & \textbf{99.75} & \textbf{100.00} & \textbf{100.00} & \textbf{0.00} \\
 & Reset & 97.34 & 0.00 & \textbf{100.00} & \textbf{0.00} \\
 & MemStrata & 73.71 & 0.00 & 0.00 & 100.00 \\
 & MemTX & 73.35 & 0.00 & 0.00 & 100.00 \\
 & TRACE & 97.84 & 97.89 & 98.95 & 1.05 \\
\midrule
Old-Stale & Static & 99.88 & \textbf{100.00} & \textbf{100.00} & \textbf{0.00} \\
 & Restore & 97.05 & 0.00 & 0.00 & 100.00 \\
 & Reset & 98.15 & 0.00 & \textbf{100.00} & \textbf{0.00} \\
 & MemStrata & 99.88 & \textbf{100.00} & \textbf{100.00} & \textbf{0.00} \\
 & MemTX & \textbf{99.94} & \textbf{100.00} & \textbf{100.00} & \textbf{0.00} \\
 & TRACE & 98.79 & 98.79 & \textbf{100.00} & \textbf{0.00} \\
\midrule
Average & Static & 88.04 & 50.00 & 50.00 & 50.00 \\
 & Restore & \textbf{98.40} & 50.00 & 50.00 & 50.00 \\
 & Reset & 97.75 & 0.00 & \textbf{100.00} & \textbf{0.00} \\
 & MemStrata & 86.79 & 50.00 & 50.00 & 50.00 \\
 & MemTX & 86.65 & 50.00 & 50.00 & 50.00 \\
 & TRACE & 98.32 & \textbf{98.34} & 99.47 & 0.53 \\
\bottomrule
\end{tabular*}
\end{table}

\clearpage
\section{Detailed Results: DeepSeek}
\label{app:deepseek}

\subsection{Memora}
\label{app:deepseek-memora}

Table~\ref{tab:deepseek-memora} reports the three departure conditions and their
equal-weight macro-average for the DeepSeek actor. Scores are reported to two
decimal places; averaging rounded condition scores can differ in the last digit.

\begin{table}[H]
\caption{DeepSeek on Memora at three departure points (\%). Average is the equal-weight macro-average of the three conditions. Bold marks the best reported value among the six methods in each block.}
\label{tab:deepseek-memora}
\centering
\small
\setlength{\tabcolsep}{6pt}
\begin{tabular}{@{}llcccc@{}}
\toprule
Departure & Method & Overall$\uparrow$ & VIA$\uparrow$ & IIR$\uparrow$ & FAMA$\uparrow$ \\
\midrule
25\% & Static & 66.66 & 49.88 & 78.02 & 39.72 \\
 & Restore & 69.22 & 41.02 & 91.77 & 37.57 \\
 & Reset & \textbf{72.52} & 40.39 & \textbf{99.19} & 39.95 \\
 & MemStrata & 68.43 & 51.78 & 79.54 & 42.24 \\
 & MemTX & 68.69 & \textbf{52.39} & 79.64 & \textbf{42.66} \\
 & TRACE & 67.83 & 50.71 & 79.67 & 40.77 \\
\midrule
50\% & Static & 67.51 & 51.99 & 77.79 & 41.50 \\
 & Restore & 69.50 & 42.30 & 90.62 & 38.45 \\
 & Reset & \textbf{72.79} & 39.98 & \textbf{99.59} & 39.82 \\
 & MemStrata & 68.36 & \textbf{53.02} & 78.15 & 42.35 \\
 & MemTX & 69.69 & 52.86 & 80.63 & \textbf{43.39} \\
 & TRACE & 68.06 & 50.88 & 79.15 & 40.99 \\
\midrule
75\% & Static & 70.93 & 55.53 & 84.81 & 47.30 \\
 & Restore & 70.71 & 43.48 & 91.00 & 39.79 \\
 & Reset & \textbf{72.21} & 39.53 & \textbf{99.04} & 39.13 \\
 & MemStrata & 71.15 & 55.89 & 82.33 & 47.39 \\
 & MemTX & 71.28 & \textbf{56.28} & 82.75 & \textbf{48.12} \\
 & TRACE & 70.61 & 54.00 & 83.33 & 45.99 \\
\midrule
Average & Static & 68.36 & 52.47 & 80.20 & 42.84 \\
 & Restore & 69.81 & 42.27 & 91.13 & 38.60 \\
 & Reset & \textbf{72.51} & 39.97 & \textbf{99.27} & 39.63 \\
 & MemStrata & 69.31 & 53.57 & 80.01 & 43.99 \\
 & MemTX & 69.89 & \textbf{53.84} & 81.01 & \textbf{44.72} \\
 & TRACE & 68.83 & 51.86 & 80.72 & 42.58 \\
\bottomrule
\end{tabular}
\end{table}

\clearpage
\subsection{STALE Type~II}
\label{app:deepseek-stale}

Table~\ref{tab:deepseek-stale} separates the Early, Middle, and Late buckets
and reports their equal-weight macro-average. Table~\ref{tab:main-results}
uses the Overall, VIA, and IIR entries from this Average block. The DeepSeek
aggregate weights buckets equally despite their different sizes; the
Qwen aggregate instead pools episodes. Cross-model score differences therefore
may also reflect the aggregation rule.

\begin{table}[H]
\caption{DeepSeek on STALE Type~II (\%). Early, Middle, and Late contain 64, 64, and 72 episodes, respectively. Average gives equal weight to the three buckets. MemStrata and MemTX results for Early, Middle, and Late buckets are omitted due to data inconsistencies; only Average results are reported. Bold marks the best reported value among the six methods in each block.}
\label{tab:deepseek-stale}
\centering
\small
\setlength{\tabcolsep}{3pt}
\begin{tabular*}{\linewidth}{@{\extracolsep{\fill}}llcccccc@{}}
\toprule
Bucket & Method & Overall$\uparrow$ & Dim1$\uparrow$ & Dim2$\uparrow$ & VIA / Dim3$\uparrow$ & IIR$\uparrow$ & LS$\uparrow$ \\
\midrule
Early & Static & 25.00 & 17.19 & 17.19 & 40.63 & 12.50 & 12.50 \\
 & Restore & 6.25 & 0.00 & 3.13 & 15.63 & 0.00 & 0.00 \\
 & Reset & 13.02 & 0.00 & 14.06 & 25.00 & 0.00 & 0.00 \\
 & TRACE & \textbf{71.88} & \textbf{76.56} & \textbf{70.31} & \textbf{68.75} & \textbf{70.31} & \textbf{59.38} \\
\midrule
Middle & Static & 24.48 & 20.31 & 20.31 & 32.81 & 15.63 & 9.38 \\
 & Restore & 6.25 & 0.00 & 1.56 & 17.19 & 0.00 & 0.00 \\
 & Reset & 17.19 & 0.00 & 28.13 & 23.44 & 0.00 & 0.00 \\
 & TRACE & \textbf{46.88} & \textbf{43.75} & \textbf{42.19} & \textbf{54.69} & \textbf{40.63} & \textbf{37.50} \\
\midrule
Late & Static & 16.67 & 9.72 & 11.11 & 29.17 & 9.72 & 6.94 \\
 & Restore & 5.56 & 1.39 & 0.00 & 15.28 & 0.00 & 0.00 \\
 & Reset & 18.06 & 0.00 & 31.94 & 22.22 & 0.00 & 0.00 \\
 & TRACE & \textbf{44.44} & \textbf{41.67} & \textbf{41.67} & \textbf{50.00} & \textbf{38.89} & \textbf{31.94} \\
\midrule
Average & Static & 22.05 & 15.74 & 16.20 & 34.20 & 12.62 & 9.61 \\
 & Restore & 6.02 & 0.46 & 1.56 & 16.03 & 0.00 & 0.00 \\
 & Reset & 16.09 & 0.00 & 24.71 & 23.55 & 0.00 & 0.00 \\
 & MemStrata & 26.00 & 16.15 & 25.64 & 36.23 & 14.58 & 11.63 \\
 & MemTX & 26.89 & 19.21 & 24.71 & 36.75 & 14.58 & 10.07 \\
 & TRACE & \textbf{54.40} & \textbf{53.99} & \textbf{51.39} & \textbf{57.81} & \textbf{49.94} & \textbf{42.94} \\
\bottomrule
\end{tabular*}
\end{table}

\clearpage

\subsection{ManBench-Return}
\label{app:deepseek-manbench}

Table~\ref{tab:deepseek-manbench} reports both conditions and their
equal-weight average for all six methods. The condition definitions match the Qwen
evaluation in Appendix~\ref{app:qwen-manbench}. WSAR measures admission of
incorrect state to the return context.

\begin{table}[H]
\caption{DeepSeek on ManBench-Return (\%). Average gives equal weight to Old-Valid and Old-Stale using unrounded condition scores; displayed condition values may not reproduce the final decimal. Overall measures final answer correctness. Bold marks the best reported value among the six methods in each block.}
\label{tab:deepseek-manbench}
\centering
\small
\setlength{\tabcolsep}{6pt}
\begin{tabular}{@{}llcccc@{}}
\toprule
Condition & Method & Overall$\uparrow$ & VIA$\uparrow$ & IIR$\uparrow$ & WSAR$\downarrow$ \\
\midrule
Old-Valid & Static & 31.06 & 0.00 & 0.00 & 100.00 \\
 & Restore & \textbf{99.88} & \textbf{100.00} & \textbf{100.00} & \textbf{0.00} \\
 & Reset & 95.39 & 0.00 & \textbf{100.00} & \textbf{0.00} \\
 & MemStrata & 52.50 & 0.00 & 0.00 & 100.00 \\
 & MemTX & 52.38 & 0.00 & 0.00 & 100.00 \\
 & TRACE & 92.04 & 91.86 & 96.89 & 3.11 \\
\midrule
Old-Stale & Static & 99.85 & \textbf{100.00} & \textbf{100.00} & \textbf{0.00} \\
 & Restore & 87.13 & 0.00 & 0.00 & 100.00 \\
 & Reset & 97.38 & 0.00 & \textbf{100.00} & \textbf{0.00} \\
 & MemStrata & 99.85 & \textbf{100.00} & \textbf{100.00} & \textbf{0.00} \\
 & MemTX & \textbf{100.00} & \textbf{100.00} & \textbf{100.00} & \textbf{0.00} \\
 & TRACE & 93.22 & 93.30 & \textbf{100.00} & \textbf{0.00} \\
\midrule
Average & Static & 65.45 & 50.00 & 50.00 & 50.00 \\
 & Restore & 93.51 & 50.00 & 50.00 & 50.00 \\
 & Reset & \textbf{96.39} & 0.00 & \textbf{100.00} & \textbf{0.00} \\
 & MemStrata & 76.18 & 50.00 & 50.00 & 50.00 \\
 & MemTX & 76.19 & 50.00 & 50.00 & 50.00 \\
 & TRACE & 92.63 & \textbf{92.58} & 98.44 & 1.56 \\
\bottomrule
\end{tabular}
\end{table}

\clearpage
\section{Component Ablations on STALE Type II}
\label{app:core-ablation}

This study examines four mechanisms in TRACE using Qwen on STALE Type~II.
Each variant removes one mechanism from the full admission procedure.
The abbreviations II, OAS, CVG, and SBPR are defined in
Section~\ref{sec:rq1} and identify the mechanisms removed by each variant.
Figure~\ref{fig:stale-core-ablation} presents Overall, VIA, and IIR;
Table~\ref{tab:stale-core-ablation} also reports LS, which requires all three
probes to pass in the same episode. The metric definitions follow
Section~\ref{sec:metrics}.

\begin{table}[htbp]
  \centering
  \small
  \setlength{\tabcolsep}{10pt}
  \caption{Component ablations on STALE Type~II with Qwen (\%). Bold marks
  the highest score in each column. Each ablation removes one mechanism.}
  \label{tab:stale-core-ablation}
  \begin{tabular}{@{}lcccc@{}}
    \toprule
    Method & Overall$\uparrow$ & VIA$\uparrow$ & IIR$\uparrow$ & LS$\uparrow$ \\
    \midrule
    Full TRACE & \textbf{52.33} & \textbf{52.50} & \textbf{50.50} & \textbf{39.00} \\
    w/o II & 26.50 & 34.00 & 19.50 & 18.00 \\
    w/o OAS & 38.67 & 35.50 & 36.00 & 29.00 \\
    w/o CVG & 47.33 & 48.00 & 44.00 & 35.00 \\
    w/o SBPR & 42.00 & 43.00 & 38.50 & 31.00 \\
    \bottomrule
  \end{tabular}
\end{table}

\paragraph{w/o II.}
This variant disables cross-attribute implicit invalidation while retaining
explicit updates, revocations, and the remaining admission checks. It tests
whether later observations must be used to reconsider an earlier memory even
when they do not explicitly retract it. Its IIR decreases from 50.50\% to
19.50\%, and LS decreases from 39.00\% to 18.00\%. This is the largest
decline among the four ablations on both metrics.

\paragraph{w/o OAS.}
This variant replaces obligation-aware evidence selection with relevance
ranking under the same context budget. The final coverage check remains
active. It tests whether selecting individually relevant records supplies the
evidence needed for the full returning workload. Overall falls from 52.33\%
to 38.67\%, while VIA falls from 52.50\% to 35.50\%.

\paragraph{w/o CVG.}
This variant retains the evidence selection procedure but allows the
returning workload to proceed when critical-obligation coverage is incomplete.
Identity, provenance, and budget checks remain active. It isolates the
decision to withhold an incomplete view from the preceding selection step.
Overall decreases by 5.00 percentage points and LS by 4.00 percentage points.
These aggregate results support the gate's contribution in this study; they
do not imply that every rejected view would otherwise yield an incorrect
answer.

\paragraph{w/o SBPR.}
This variant leaves the public Return View unchanged but removes the
requirement that retrieved private experiences be bound to admitted sources.
Owner isolation remains active. It tests whether information excluded from
the public view can affect the answer through private-memory retrieval.
Overall falls to 42.00\%, IIR to 38.50\%, and LS to 31.00\%. The reduction
is consistent with the need to apply admission constraints to both public
work state and private experiences.

\clearpage
\section{Agent-Scale and Backend Studies}
\label{app:scale-backend}

These studies examine two distinct questions: whether return-time admission
remains stable as the team grows, and whether the same admission boundary can
operate above different memory representations. The first varies the number
of active agents in ManBench-Return; the second varies the memory backend in
Memora while holding the return condition fixed.

\subsection{Agent-scale diagnostic}
\label{app:agent-scale-study}

We vary the number of active agents over 4, 8, 16, and 32 while keeping
one returning agent. Static and TRACE use the same balanced Old-Valid/Old-Stale
construction at each scale. Overall measures final-answer correctness;
VIA and IIR distinguish preservation of valid state from rejection of invalid
state. Table~\ref{tab:agent-scale} gives the scale-specific results, and
Table~\ref{tab:agent-scale-average} summarizes them with equal weight per scale.

\begin{table}[htbp]
\centering
\caption{Macro-average across the four agent scales in
Table~\ref{tab:agent-scale} (\%). WSAR is the complement of IIR in this
admission diagnostic. Bold marks the better value in each column.}
\label{tab:agent-scale-average}
\small
\setlength{\tabcolsep}{12pt}
\begin{tabular}{lcccc}
\toprule
Method & Overall$\uparrow$ & VIA$\uparrow$ & IIR$\uparrow$ & WSAR$\downarrow$ \\
\midrule
Static & 76.51 & 44.24 & 44.24 & 55.76 \\
TRACE & \textbf{96.07} & \textbf{96.76} & \textbf{98.83} & \textbf{1.17} \\
\bottomrule
\end{tabular}
\end{table}

TRACE's Overall ranges from 95.85\% to 96.21\%, VIA from 96.62\% to
96.92\%, and IIR from 98.72\% to 98.94\%. The corresponding macro-average
gains over Static are 19.56, 52.52, and 54.59 percentage points. These results
support stability within the evaluated range, without a monotonic increase
in performance or advantage as the team grows. This full-coverage diagnostic
is separate from the frozen five-agent comparison in
Table~\ref{tab:main-results}.

\subsection{Backend integration and paired evaluation}
\label{app:backend-study}

\paragraph{What changes with the backend.}
ScopedMem is our agent-scoped versioned reference store. Mem0 extracts,
consolidates, and retrieves conversational memories~\citep{chhikara2025mem0};
Memobase organizes user profiles and events;
A-MEM constructs linked notes whose attributes and connections can evolve
as new memories arrive~\citep{xu2025amem}. The adapters preserve these native
formation and retrieval operations. TRACE controls eligibility at return
through a common representation of record identity, text, retrieval score,
and source bindings. It does not replace each backend with a shared store
or use reference-store text as a fallback for native retrieval.

\paragraph{Admission across different representations.}
Backend substitution requires more than a common search API: a retrieved
profile or linked note may combine several historical inputs. The adapter
must retain the evidence associations needed to decide whether that memory
can enter the returning agent's context. A source association identifies
origin; it does not establish current validity. Where the native system
does not expose exact dependencies, conservative source associations can
restrict retention. Thus the granularity of source attribution is an
integration constraint, and native retrieval relevance cannot substitute
for TRACE's validity decision. Appendix~\ref{app:method-details} details
the source-bound readmission rule.

\paragraph{Paired protocol.}
The study uses the 137 Memora episodes at 50\% departure with Qwen3.5-122B-A10B
as the actor and \texttt{gpt-5.6-terra} as the scorer. Within each backend,
Static and TRACE share the episode history and pre-return memory formation.
Static follows the ordinary retrieval path; TRACE restricts readmission
using the admitted work state. Branch construction imports the stored
records without repeating memory formation, and retrieval remains native
to the selected backend. This comparison measures the effect of attaching
TRACE within each backend. It does not require identical retrieved text
across backends, whose memory construction and ranking differ.

\begin{table}[H]
\centering
\caption{Backend portability on Memora at 50\% departure (\%). Each
Static/TRACE pair shares its backend, actor, and scorer. Bold marks the
higher value within each pair.}
\label{tab:backend-portability}
\small
\setlength{\tabcolsep}{12pt}
\begin{tabular}{llccc}
\toprule
Backend & Method & VIA$\uparrow$ & IIR$\uparrow$ & FAMA$\uparrow$ \\
\midrule
\multirow{2}{*}{ScopedMem} & Static & \textbf{55.04} & 67.77 & 39.13 \\
& TRACE & 52.01 & \textbf{74.19} & \textbf{39.31} \\
\addlinespace[3pt]
\multirow{2}{*}{Mem0} & Static & \textbf{53.53} & 68.93 & 37.81 \\
& TRACE & 53.48 & \textbf{70.06} & \textbf{38.29} \\
\addlinespace[3pt]
\multirow{2}{*}{Memobase} & Static & \textbf{54.54} & 67.94 & \textbf{38.52} \\
& TRACE & 53.19 & \textbf{68.33} & 37.17 \\
\addlinespace[3pt]
\multirow{2}{*}{A-MEM} & Static & \textbf{55.33} & 67.24 & \textbf{39.21} \\
& TRACE & 54.00 & \textbf{68.35} & 38.09 \\
\bottomrule
\end{tabular}
\end{table}

\paragraph{Results and interpretation.}
In Table~\ref{tab:backend-portability}, TRACE increases IIR by 6.42, 1.13,
0.39, and 1.11 percentage points for ScopedMem, Mem0, Memobase, and A-MEM,
respectively. VIA decreases by 3.03, 0.05, 1.35, and 1.33 percentage points.
FAMA changes by $+0.18$, $+0.48$, $-1.35$, and $-1.12$ percentage points:
greater rejection of invalid information does not consistently improve
the combined remembering-and-forgetting score. The paired results support
the feasibility of applying return-time admission above all four tested
backends, while showing a retention--rejection trade-off on Memora's explicit
updates. They do not isolate which representation or source-binding choice
causes a particular gain or loss. The results also do not establish
compatibility with arbitrary stores, or comparative backend latency and
token efficiency.

\end{document}